\documentclass{aa} 
\usepackage{float} 
\usepackage{graphicx}  
\usepackage{graphics}
\usepackage{txfonts}  
\usepackage{natbib}
\usepackage{adjustbox}
\bibpunct{(}{)}{;}{a}{}{,}
\usepackage[dvipsnames]{xcolor}
\usepackage[normalem]{ulem}

\begin{document}  

% \bibpunct{(}{)}{;}{a}{}{,} % to follow the A&A style 
 
\title {Fundamental properties of a selected sample of Ap stars: Inferences from interferometric and asteroseismic constraints}

\author{M. Deal\inst{1} \and M. S. Cunha\inst{1}\and Z. Keszthelyi\inst{2} \and K. Perraut\inst{3} \and D. L. Holdsworth\inst{4}}
  
\institute{Instituto de Astrofísica e Ciências do Espaço, Universidade do Porto
CAUP, Rua das Estrelas, PT4150-762 Porto, Portugal     
           \and
Anton Pannekoek Institute for Astronomy, University of Amsterdam, Science Park 904, 1098 XH, Amsterdam, The Netherlands 
           \and
           Univ. Grenoble Alpes, CNRS, IPAG, 38000 Grenoble, France
           \and
           Jeremiah Horrocks Institute, University of Central Lancashire, Preston PR1 2HE, UK
\\
            \email{morgan.deal@astro.up.pt} 
           }
           
\date{\today}

\abstract
% Context
{Magnetic fields influence the formation and evolution of stars and impact the observed stellar properties. Ap stars (magnetic A-type stars) are a prime example of this. Access to precise and accurate determinations of their stellar fundamental properties, such as masses and ages, is crucial to understand the origin and evolution of fossil magnetic fields.}
% Aims
{We propose using the radii and luminosities determined from interferometric measurements, in addition to seismic constraints when available, to infer fundamental properties of 14 Ap stars préviously characterised.}
% Method
{We used a grid-based modelling approach, employing stellar models computed with the \textsc{cestam} stellar evolution code, and the parameter search performed with the \textsc{aims} optimisation method. The stellar model grid was built using a wide range of initial helium abundances and metallicities in order to avoid any bias originating from the initial chemical composition. The large frequency separations ($\Delta\nu$) of HR\,1217 (HD\,24712) and $\alpha$~Cir (HD\,128898), two rapidly oscillating Ap stars of the sample, were used as seismic constraints.}
% Results
{We inferred the fundamental properties of the 14 stars in the sample. The overall results are consistent within $1\sigma$ with previous studies, however, the stellar masses inferred in this study are higher. This trend likely originates from the broader range of chemical compositions considered in this work. We show that the use of $\Delta\nu$ in the modelling significantly improves our inferences, allowing us to set reasonable constraints on the initial metallicity which is, otherwise, unconstrained. This gives an indication of the efficiency of atomic diffusion in the atmospheres of roAp stars and opens the possibility of characterising the transport of chemical elements in their interiors.}
% Conclusions
{}

  
\titlerunning{Fundamental properties of a selected sample of Ap stars}
  
\authorrunning{}  

\maketitle 

%_____________________________________________________________________
%%%%%%%%%%%%%%%%%%%%%%%%%%%%%%%%%%%%%%%%%%%%%%%%%%%%%%%%%%%%%%%%%%%
\section{Introduction}

Magnetism is ubiquitously present on various scales in the Universe. Magnetic fields have a significant impact on star formation \citep[e.g.][]{commercon11,mackey11}, as well as stellar structure and evolution \citep[e.g.][]{mestel99,donati09}. The first stellar magnetic field of a star other than the Sun was detected by \cite{babcock47}. This star, 78~Vir, is a chemically-peculiar A-type star (Ap star). Consequently, research of Ap stars has a long and rich history \citep[e.g.][]{babcock58,wolff67,wolff68,landstreet82,auriere07,sikora19}. They amount to only a few per cent of the A-type-star population \citep{sikora19}, yet their observation and modelling has a significant scientific potential. Ap star characterisation, both on a star-by-star basis and in terms of their ensemble properties, provides clues as to the origin of strong, large-scale, fossil stellar magnetic fields \citep[e.g.][and references therein]{cowling45,braithwaite17,cantiello19}, and can contribute to our understanding of how these fields influence both stellar evolution \citep[e.g.][]{Keszthelyi19,schneider20} and the physical processes leading to the segregation of chemical elements via atomic diffusion \cite[e.g.][]{michaud15,kochukhov18}. Unfortunately, despite their scientific interest, Ap stars are not easy to characterise. In fact, classical stellar parameters of Ap stars derived from the analysis of photometric and/or spectroscopic data can be biased due to the surface chemical peculiarities. This, in turn, can lead to inaccurate determinations of the stars' fundamental properties, such as the mass, radius, and age, compromising tests to our theoretical understanding of how these stars evolve and become chemically peculiar. In this context, the study of stars for which one may hope to derive unbiased classical parameters and fundamental properties, becomes particularly relevant. An example of these types of benchmarks are stars whose angular diameter can be directly measured through interferometry. With this in mind, over the past years an effort has been made to observe all Ap stars within reach of currently available interferometric instruments in terms of sensitivity and angular resolution. Ap stars having angular diameters smaller than 1 millisecond of arc requires operating in the visible range and with hectometric baselines. As a consequence, most of our targets were observed in the northern hemisphere with the CHARA array \citep{bruntt08,bruntt10,Romanovskaya19,perraut11,perraut13,perraut15,perraut16,perraut20}. Together with state-of-the-art parallaxes and bolometric fluxes, the measured angular diameters were used to infer nearly model-independent radii, luminosities, and effective temperatures for this sample of benchmark Ap stars. The properties of the full sample, composed of 14 Ap stars, have recently been discussed by \cite{perraut20}.

Among the Ap stars, there is a subgroup known as the rapidly oscillating Ap stars (hereafter, roAp stars), which exhibit high frequency pulsations \citep{kurtz82}. Their effective temperatures range from about 6000 to 9000 K and their pulsation periods from about 5~min \citep{cunha19} to 24~min \citep{alentievetal12}. Up to now, most roAp stars have been discovered through the analysis of ground-based photometric time series, but the NASA TESS (Transiting Exoplanet Survey Satellite) is rapidly increasing the number of new detections \citep[e.g.][]{cunha19,balona19}. The combination of interferometric and asteroseimic data has significant constraining power in the context of stellar modelling \citep{creevey07,cunha07}, making roAp stars with a measured angular diameter primary targets for modelling.

The aim of the present paper is to use the interferometric radii and luminosities of the 14 stars characterised by \cite{perraut20} to infer their masses and ages. Out of these 14 stars, five are roAp stars. The inference is performed with the grid-based optimisation method \textsc{aims} (Asteroseismic Inference on a Massive Scale, \citealt{rendle19}). The stellar models are computed with the \textsc{cestam} evolution code (Code d'Evolution Stellaire Adaptatif et Modulaire, the 'T' stands for transport). To perform inferences as unbiased as possible, the grid includes various initial chemical compositions. In addition to precise measurements of radii and luminosities, we assess the benefit of having seismic constraints, when available. In order to further strengthen the robustness of these results, we additionally use \textsc{mesa} software \citep{paxton11,paxton13,paxton15,paxton18,paxton19} where the effects of surface fossil magnetic fields have previously been implemented by \cite{Keszthelyi19,keszthelyi20}.

In Section 2, we present the Ap stars sample and the information provided by roAp pulsations to constrain stellar models. The stellar models used to infer stellar parameters are described in Section 3 and the optimisation procedure in Section 4. We present the inferred masses, ages and hydrogen mass fraction in the core ($X_c$) using classical and seismic constraints in Section 5. We address the impact of neglecting the magnetic field and transport processes of chemical elements in Section 6. We finally discuss the results in Section~7 and conclude in Section~8.

\section{Ap stars sample}

\subsection{Properties of the sample}
\begin{figure}[!ht]
\center
\includegraphics[scale=0.58]{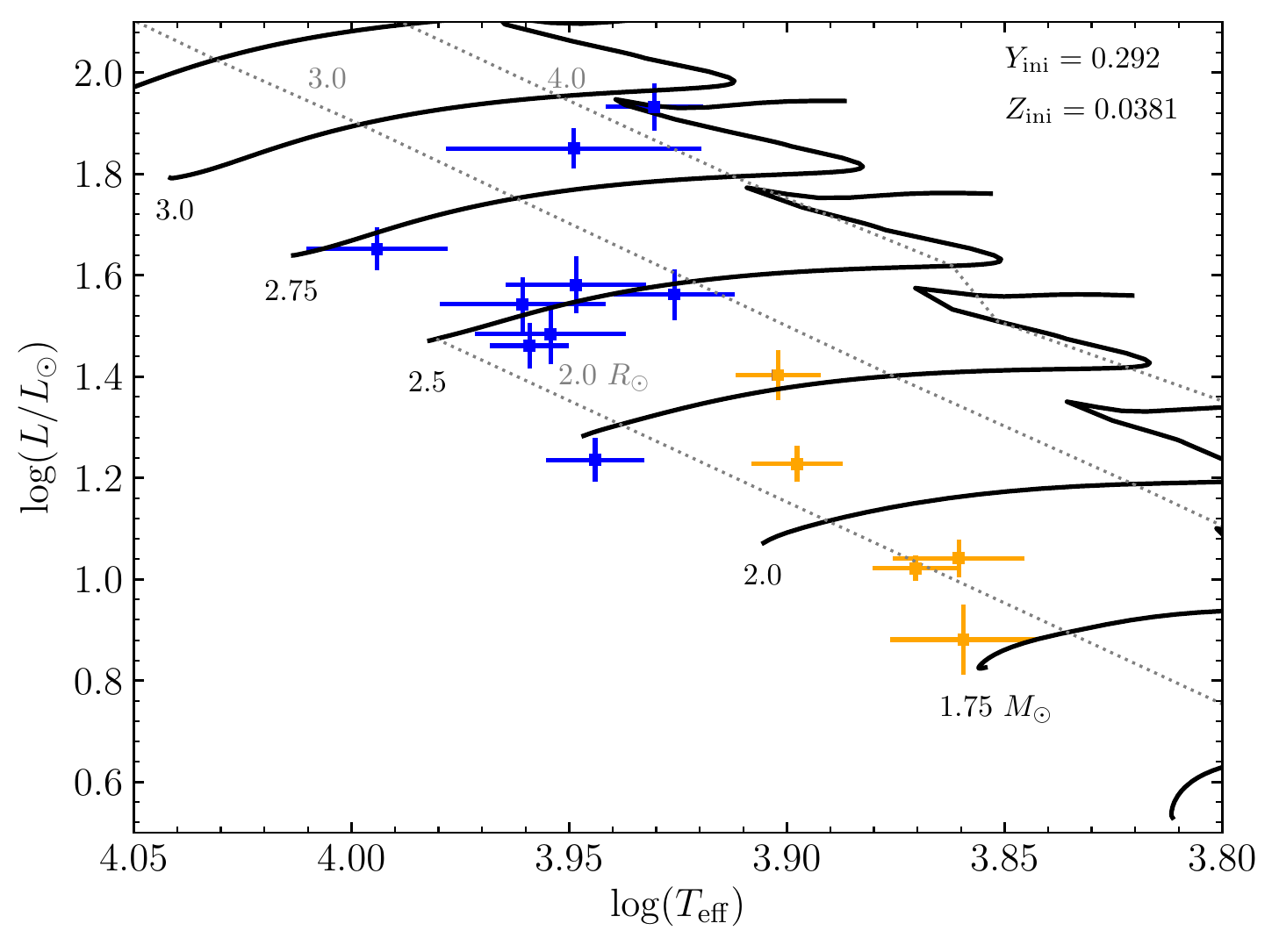}
\includegraphics[scale=0.58]{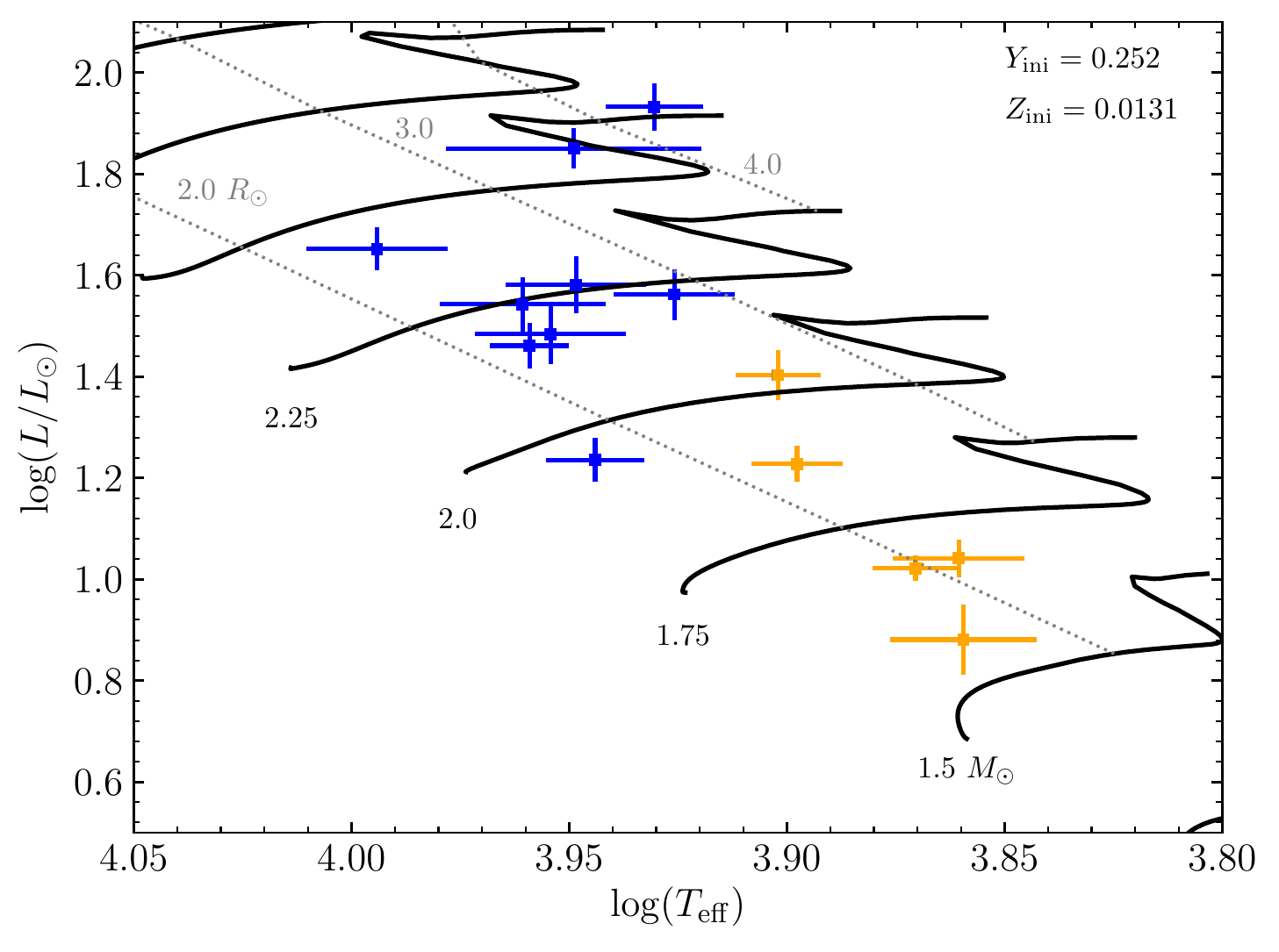}
\includegraphics[scale=0.58]{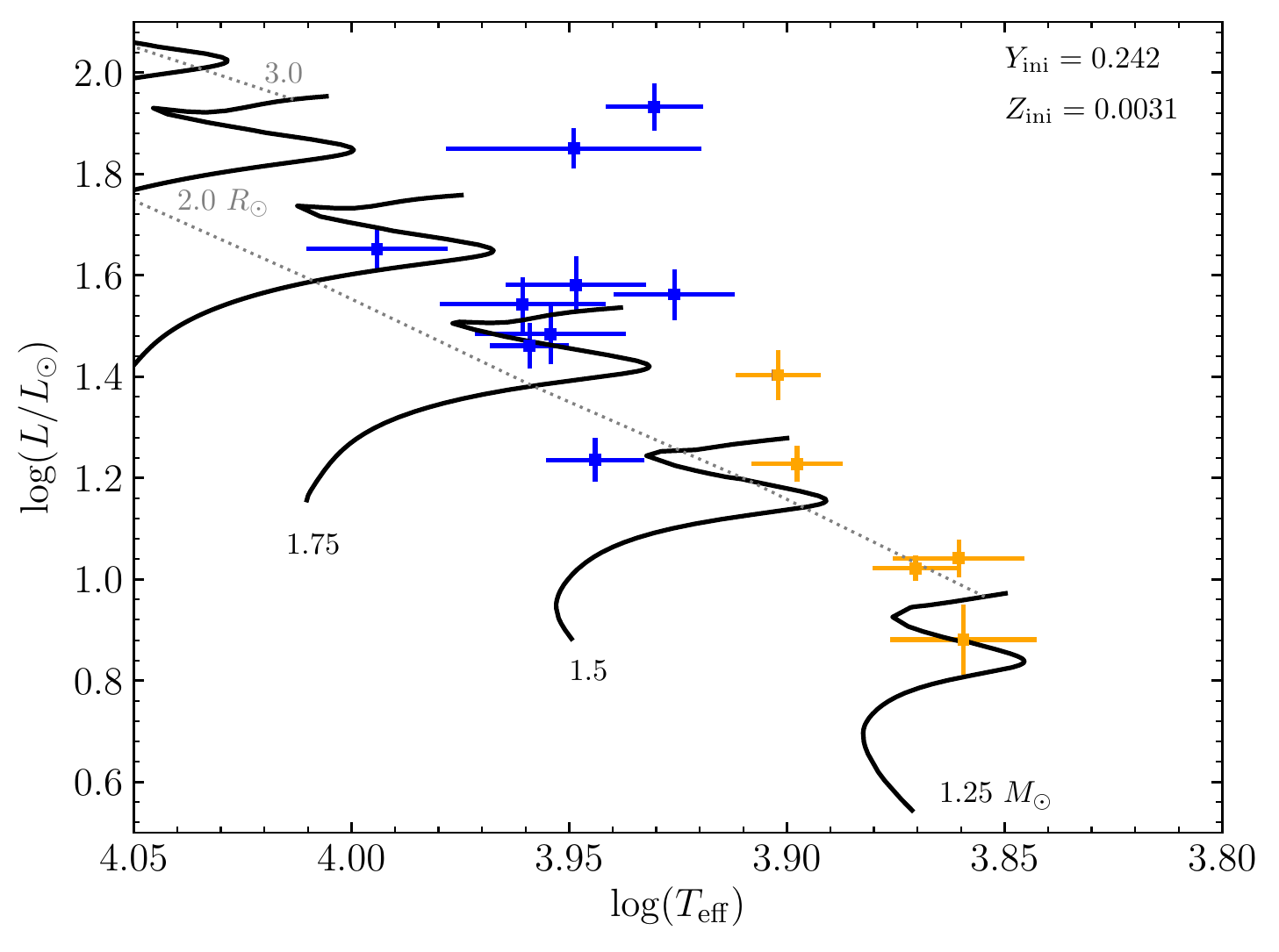}
\caption{HR diagram including the studied stars listed in Table \ref{table:1} and evolutionary tracks for masses between $1.25$ and $3.25~$M$_\odot$ at different initial chemical composition (black solid lines). Orange crosses are roAp stars while the blue ones are the other star of the sample. The dotted lines represent the iso-radius at $2.0$, $3.0$, and $4.0$~R$_\odot$. See the text for details about the models.}
\label{fig:HR}
\end{figure}

This study focuses on the stars characterised by \cite{perraut20}. The sample is composed of 14 Ap stars with angular diameters measured through interferometry (five of them are roAp stars). They were chosen for the interferometry programme because they are brighter than the current sensitivity limit of visible interferometers (about R~=~7 in standard atmospheric conditions to derive accurate angular diameters, \citealt{perraut20}), and have estimated angular diameters greater than about 0.2~mas, thus within the current angular resolution of these instruments. Information on the individual stars is given in Table~\ref{table:1}, where the effective temperature, luminosity, and radius were extracted from table~4 of the paper by \cite{perraut20}. For each star, the authors derived these properties from the angular diameter, bolometric flux, and parallax. The adopted parallaxes were retrieved from the GAIA DR2 \citep{gaiaDR2} for all stars except for the two brightest targets, namely, $\alpha$~Cir (HD\,128898) and HD\,137909 (the latter being a binary) for which the authors adopted, instead, the parallaxes from Hipparcos~\citep{Hipparcos}. The bolometric flux was computed from the observed spectral energy distribution obtained by combining photometric and flux-calibrated spectroscopic data at different wavelengths. Figure \ref{fig:HR} shows the stars in an Hertzsprung-Russell diagram along with evolutionary tracks and iso-radii, for different initial chemical compositions. Models are computed with the \textsc{cestam} evolution code using the input physics as described in Section~\ref{models}. A core overshoot with a step extend of 0.15~Hp is adopted for these tracks.

\begin{table*}
\centering
\caption{Properties of the Ap star sample. Radii, luminosities, effective temperatures, and average surface magnetic field strengths are taken from \cite{perraut20} and references therein except where specified otherwise. The large frequency separations are derived from the literature following the description given in the text (see Sections~\ref{seismology} and \ref{sismo}).} 
\label{table:1}
\begin{adjustbox}{width=18.5cm,center}
\begin{tabular}{llccccccccc}
\noalign{\smallskip}\hline\hline\noalign{\smallskip}
HD  & Other name & Type & $T_\mathrm{eff}$ [K] &  $L$ [$L_\odot$] & $R$ [$R_\odot$] & $P_\mathrm{rot}$ [d] & $<B_{\rm s}>$ [kG] &$\Delta\nu$ [$\mu\mathrm{Hz}$] & $\Delta\nu_\mathrm{mag}$ [$\mu\mathrm{Hz}$]   \\
\noalign{\smallskip}\hline\noalign{\smallskip}
4778 & GO And & noAp\tablefoottext{1} & $9135\pm400$ & $34.9\pm4.3$ & $2.36\pm0.12$ & 2.56\tablefoottext{b}& 2.6 &- &- \\
24712 & HR\,1217 & roAp\tablefoottext{2} & $7235\pm280$ & $7.6\pm1.2$ & $1.75\pm0.05$ & 12.46\tablefoottext{a} & 2.3\tablefoottext{a} & $67.76\pm0.13$ & [$64.8$, $68.8$] \\
108662 & 17 Com A & noAp\tablefoottext{1} & $8880\pm330$ & $38.1\pm4.9$ & $2.59\pm0.12$ & 5.08\tablefoottext{a} & 3.3\tablefoottext{d} &-& - \\
108945 & 21 Com   & noAp\tablefoottext{1,3} & $8430\pm270$ & $36.5\pm4.2$ & $2.82\pm0.09$ & 2.05\tablefoottext{a} & 0.6\tablefoottext{a} &-& - \\
118022 & 78 Vir   & noAp\tablefoottext{1,4} & $9100\pm190$ & $28.9\pm3.0$ & $2.17\pm0.06$ & 3.72\tablefoottext{a} & 3.0\tablefoottext{e} &-& - \\
120198 &  84 UMa  & noAp\tablefoottext{1,5} & $9865\pm370$ & $44.9\pm4.3$ & $2.28\pm0.10$ & 1.39\tablefoottext{a} & 1.1\tablefoottext{a} &-& - \\
128898 & $\alpha$~Cir & roAp\tablefoottext{6} & $7420\pm170$ & $10.51\pm0.60$ & $1.967\pm0.066$ & 4.48\tablefoottext{a} & 1.0\tablefoottext{a} & $60.37\pm0.03$ & [$57.4$, $61.4$] \\
137909& $\beta$~CrB & roAp\tablefoottext{7} & $7980\pm180$ & $25.3\pm2.9$ & $2.63\pm0.09$ & 18.49\tablefoottext{a} & 5.5\tablefoottext{f} &-& - \\
153882 & V451 Her & noAp\tablefoottext{1} & $8980\pm600$ & $70.8\pm6.5$ & $3.46\pm0.37$ & 6.01\tablefoottext{b}& 3.8 & - \\
176232 & 10~Aql & roAp\tablefoottext{8} & $7900\pm190$ & $16.9\pm1.4$ & $2.21\pm0.08$ & 6.05\tablefoottext{b}& 1.5 &-& - \\
188041 & V1291 Aql & noAp\tablefoottext{5} & $9000\pm360$ & $30.5\pm4.2$ & $2.26\pm0.05$ & 224.50\tablefoottext{b} & 3.6 &-& - \\
201601 & $\gamma$ Equ A & roAp\tablefoottext{9}& $7253\pm235$ & $11.0\pm0.93$ & $2.11\pm0.07$ & 35462.5\tablefoottext{a} & 3.9\tablefoottext{f} &-& - \\
204411 & HR8216 & noAp\tablefoottext{1,5} & $8520\pm220$ & $85.6\pm9.2$ & $4.23\pm0.11$ & unknown\tablefoottext{c}& <0.8 &-& - \\
220825 & $\kappa$ Psc & noAp\tablefoottext{5} & $8790\pm230$ & $17.2\pm1.7$ & $1.78\pm0.03$ & 1.42\tablefoottext{a} & 1.2\tablefoottext{a} &-& - \\
\noalign{\smallskip}\hline\noalign{\smallskip}
\end{tabular}
\end{adjustbox}
\tablefoot{\tablefoottext{1}{We searched for pulsations in TESS data (see Section~\ref{seismology}).}\tablefoottext{2}{\cite{kurtz81HR1217}.}\tablefoottext{3}{e.g. \cite{kreidl90}.}\tablefoottext{4}{\cite{paunzen18}.}\tablefoottext{5}{\cite{nelson93}.}\tablefoottext{6}{\cite{kurtz81alphacir}.}\tablefoottext{7}{\cite{hatzes04}.}\tablefoottext{8}{\cite{heller88}.}\tablefoottext{9}{\cite{kurtz83}.}\tablefoottext{a}{\citealt[][and references therein]{sikora19}. The values listed correspond to $0.69\times B_d$, $B_d$ being the dipolar field}\tablefoottext{b}{\citealt[][and references therein]{netopil17}.}\tablefoottext{c}{No rotation period listed in the literature. Furthermore, 54 days of TESS data do not show any rotational variability (see Section~\ref{seismology}).}\tablefoottext{d}{\cite{romanovskaya20}.}\tablefoottext{e}{\cite{ryabchikova17}.}\tablefoottext{e}{\cite{mathys17}.}}
\end{table*}

\subsection{Seismology of Ap stars}
\label{seismology}
The roAp stars have an added valuable scientific potential, as their pulsations may provide additional constraints to stellar modelling. Unfortunately, the strong magnetic field permeating the roAp stars impacts both directly and indirectly on the pulsations, perturbing their frequencies with respect to those expected in non-magnetic stars. The direct effect follows from the distortion of the pulsations by the Lorentz force. Studies of this direct interaction have shown that frequencies can be perturbed by a significant fraction of the typical frequency separation between consecutive modes \citep{cunha00,cunha06,bigotetal00,saio04}. While the results from these theoretical studies agree qualitatively, quantitatively they show significant differences in what concerns the perturbation to individual modes, bringing into question the use of the magnetic models in direct comparisons with the individual frequencies observed in roAp stars. Nevertheless, the scenario is different when considering the large frequency separations, $\Delta\nu$, between consecutive modes of the same degree, $l$. With the exception of modes experiencing a very strong coupling with the magnetic field, the frequency perturbation is not expected to vary significantly from mode to mode: It generally increases slightly with frequency, with a mode-to-mode variation which typically does not exceed 2~$\mu$Hz, occasionally decreasing by a fraction of a $\mu$Hz. In addition to the direct effect, the magnetic field can have an indirect effect on pulsations, resulting from its impact on the equilibrium structure. In fact, it has been argued that the magnetic field may suppress subsurface convection and influence the transport of chemical elements \citep{balmforthetal01,cunha02,theado05}. While the impact from these effects on the individual frequencies can reach a significant fraction of the large frequency separation \citep{balmforthetal01}, as structural effects their impact does not change significantly from frequency to frequency, leading to a perturbation to the large frequency separation that is typically smaller than the perturbation resulting from the direct effect of the magnetic field. As a consequence, the mode frequencies in roAp stars still tend to follow a regular pattern, often being almost equally spaced in the power spectrum. When that is the case, the observed large frequency separation can be determined and used as an extra constraint in the modelling of the star based on non-magnetic models, so far as a small magnetic correction is subtracted from the observed value. 

Among the 14 stars in our sample, five were known to be roAp stars prior to the launch of TESS (identified  in Table~\ref{table:1}).
We searched the 2-min cadence TESS observations of the stars in our sample in search for additional roAp pulsators. Of the 14 stars analysed in this work, six stars (HD\,137909, HD\,153882, HD\,176232, HD\,188041, HD\,201601, and HD\,220825) do not yet have TESS photometric observations. For the eight observed stars (from which two were previously known to be roAp stars), we analysed both the Simple Aperture Photometry (SAP) and the Pre-search Data Conditioning SAP (PDC\_SAP) data to check for any signal injected to the data by the PDC pipeline.
To search for rotational variations, we calculated a Fourier spectrum from 0-10\,d$^{-1}$ (0-0.12\,mHz). In the search for pulsational variability, we iteratively pre-whitened the light curve in the low-frequency range (0-10\,d$^{-1}$; 0-0.12\,mHz) to the noise level at high frequency to remove any rotation signal and instrumental artefacts. This serves to make the noise characteristics white in the region where pulsations are usually found. Of the eight stars with TESS data, we detected pulsational variability in two previously known roAp stars (HD\,1217/HD\,24712 and $\alpha$~Cir/HD\,128898) and rotational variability in seven stars (HD\,204411 shows no variability). Thus, the available TESS data available so far does not increase the number of known roAp stars in our sample. 
 
All data considered, we find that only two roAp stars, namely HR\,1217 (HD\,24712) and $\alpha$~Cir (HD\,128898), show a regular frequency pattern. For these two stars, we shall use the observed large frequency separation as an additional constraint to the modelling. Since the large frequency separation scales with the square root of the mean density, we expect this extra constraint, together with the interferometric radius, to improve our inference of the stellar mass. 

\subsubsection{HR\,1217 (HD\,24712)}
HR\,1217 (HD\,24712) was discovered to be a roAp star by \cite{kurtz81HR1217}. Its pulsation spectrum is well characterised from long photometric time series acquired both from the ground \citep[e.g.][]{kurtz05} and space \citep{balona19}. It shows a series of equally-spaced modes, along with at least one frequency that does not follow the same pattern, the latter having been interpreted as an example of a mode that is strongly coupled with the magnetic field \citep{cunha01,kurtz02}. There is strong evidence that the equally spaced modes are of alternating even and odd degree, from which it is deduced that the large frequency separation corresponds to twice the separation between adjacent modes. The frequency separation between the main oscillation modes of HR\,1217 (HD\,24712) is very similar in the works by \cite{kurtz05} and \cite{balona19}.  Nevertheless, the values published in these works still differ by more than the formal error that would be derived from each of them. Therefore, for our modelling we combine the results from the two publications as follows: For each of them, we compute three estimates of $\Delta\nu$ from the frequencies $\nu_1$ to $\nu_5$ (where the notation follows that used in both works), by considering the combinations $\nu_3-\nu_1$, $\nu_4-\nu_2$, and  $\nu_5-\nu_3$. We then compute the average and the square root of the variance of the six estimates and use them as estimates for the value and the uncertainty, finding $\Delta\nu=67.76\pm 0.13\,\mu$Hz. In addition to this pure observational estimate of $\Delta\nu$, we consider a more conservative scenario that takes into account the potential impact of the magnetic field on  the oscillations. From the discussion in Section~\ref{seismology}, we know that our non-magnetic models may systematically underestimate the true $\Delta\nu$ by up to about 2~$\mu$Hz and occasionally overestimate it by up to a fraction of a $\mu$Hz, as a result of the direct effect of the magnetic field. Considering in addition the potential indirect effect of the magnetic field, in this second scenario we relax the observational seismic constraint by allowing $\Delta\nu$ to vary within the interval [$\Delta\nu$-3.0;~$\Delta\nu$+1.0].

\subsubsection{$\alpha$~Cir (HD\,128898)}
$\alpha$~Cir (HD\,128898) was discovered to be a roAp star by \cite{kurtz81alphacir} and has also been a target of several ground-based and space-based campaigns. A detailed analysis of its oscillation power spectrum has been performed by \cite{bruntt09} based on space-based data collected with the star tracker on board  the Wide-field Infrared Explorer ({\it WIRE}) \citep{buzasi02} and ground-based data collected at the South African Astronomical Observatory. Recently,  new space-based data, acquired by the satellites TESS and BRITE have been analysed by \cite{weiss20}, confirming the detection of the three main frequencies reported by \cite{bruntt09} ($f_6$, $f_1$ and $f_7$, according to the notation adopted in both works).  Similarly to the case of HR\,1217 (HD\,24712), these three modes are interpreted as being of alternating even and odd degrees, implying that $\Delta\nu$ corresponds to twice the separation between adjacent modes. To estimate the value of $\Delta\nu$ and its uncertainty we proceed as before. In the present case there are only three equally spaced modes and, thus, only one estimate of $\Delta\nu$ from each publication. Moreover, in the case of \cite{weiss20} we considered only the frequencies derived from the TESS data, since the S/N in the BRITE data was significantly lower. Taking the average and the square root of the variance of the values from the two publications we find $\Delta\nu=60.37\pm 0.03\mu$Hz. Finally, as discussed for HR\,1217 (HD\,24712), in addition to this observational value, we consider a second scenario to account for the effect of the magnetic field not included in our stellar models, allowing $\Delta\nu$ to vary in the interval [$\Delta\nu$-3.0;~$\Delta\nu$+1.0]. The intervals of $\Delta\nu$  considered for the two stars with seismic constraints are summarised in Table~\ref{table:1}.

\section{Stellar models}
\label{models}
The stellar models are computed with the \textsc{cestam} evolution code (Code d'Evolution Stellaire Adaptatif et Modulaire, the 'T' stands for transport). A detailed description of the code can be found in the works of \cite{morel08}, \cite{marques13}, and \cite{deal18}. The code is able to take into account several non-standard transport processes of chemical elements (i.e. atomic diffusion including radiative accelerations) and the transport of angular momentum which may have a significant impact on the stellar property inference in non-magnetic stars \citep[e.g.][]{deal20}. Despite this, accounting for chemical element transport in Ap stars is still a challenge. 

Magnetic fields strongly impact atomic diffusion in the upper atmosphere of Ap stars \citep[][and reference theirin]{alecian02a,alecian07,stift16,alecian17}. In the stars' interiors, the direct effect of  magnetic fields on atomic diffusion is negligible \citep[e.g.][]{alecian17}. Nevertheless, it is possible that they still have an indirect effect on the transport of chemical elements \citep[e.g.][]{theado05}, through their impact on the equilibrium structure, in particular the suppression of near-surface convection discussed in Section~\ref{seismology}. Moreover, a realistic transport of chemical elements (including atomic diffusion) currently requires the addition of a parametric turbulent diffusion coefficient in order to prevent the complete depletion of helium and metals from the surface, and reproduce surface abundances of F and A-type stars \citep{richer00,richard01,michaud11,verma19b,semenova20}. Magnetic fields may also impact these competing mechanisms that are thought to be related either to mass loss or turbulence in Am stars \citep[e.g.][]{vick10,michaud11}, another type of non-magnetic chemically peculiar A-type stars.
  
In addition to the potential difficulties brought about by the magnetic field effects discussed above that are currently poorly modelled in stellar evolution codes, there are practical reasons that make accounting for chemical element transport in Ap stars particularly challenging. Models including atomic diffusion require to recompute the Rosseland mean opacity at each mesh point and each time step taking into account the abundance variations. This is possible with monochromatic opacity tables, but the computational time drastically increases. Our analysis requires a large grid of models (see Section~\ref{grid}) and such grid cannot be computed in these conditions. 

For all reasons mentioned above, as a first step we decided to not take any of the processes leading to chemical transport into account in the models computed for this study, and to not use the observed surface abundances as constraints to our modelling. However, we estimate the impact of the combined effect of atomic diffusion and an additional parametrised transport process on the modelling of HR\,1217 (HD\,24712) in Section~\ref{result-sismo}, and, more generally, on the results of this paper in Section~\ref{discu-diff}. Similarly, the effects of a magnetic field on the evolution were neglected. An assessment of the effect of this assumption on the final stellar properties is presented in Section \ref{discu-mag}.

The models are computed from the PMS (Pre-Main Sequence) to a hydrogen mass fraction in the core of $X_c=10^{-11}$, to cover a part of the sub-giant phase. We use the OPAL2005 equation of state \citep{rogers02} and the OPAL95 opacity tables \citep{iglesias96} complemented at low temperatures by the Whichita opacity data \citep{ferguson05}. Nuclear reaction rates are from the NACRE compilation \citep{angulo99} except for the $^{14}N(p,\gamma)^{15}O$ reaction, for which we use the LUNA rate \citep{imbriani04}. We use an AGSS09 initial mixture of metal \citep{asplund09} with meteoritic abundances for refractory elements \citep{serenelli10}. Convection is treated following the \cite{canuto96} formalism with a solar calibrated $\alpha_\mathrm{CGM}=0.634$. Overshoot of the convective core is taken into account with a step extend of $\alpha_\mathrm{ovs} \times \mathrm{min}(Hp,r_\mathrm{cc})$ where $r_\mathrm{cc}$ is the radius of the Schwarzschild convective core. Atmospheres are computed in the Eddington grey approximation with no mass loss taken into account. \textsc{CESTAM} follows the chemical elements individually (H, He, C, N, O, Ne, Na, Mg, Al ,Si, S, Ca, and Fe). In the following sections, [Fe/H]\footnote{[Fe/H]=$\log_{10}(N_\mathrm{Fe}/N_\mathrm{H})-\log_{10}(N_\mathrm{Fe}/N_\mathrm{H})_\odot$} refers to the iron surface abundance, and [M/H]\footnote{[M/H]=$\log_{10}(Z/X)-\log_{10}(Z/X)_\odot$, $Z$ and $X$ being the metal and hydrogen mass fraction, respectively.} to the surface metallicity. $Z$ is the mass fraction of metals. As we neglected transport of chemical elements in the models, the model values for these quantities remain unchanged with time and equal to their initial values (hereafter referred to as [Fe/H]$_\mathrm{ini}$, [M/H]$_\mathrm{ini}$ and $Z_\mathrm{ini}$). However, that it is not the case for the models considered at the end of Section \ref{result-sismo} and in Section \ref{discu-diff}, for which we consider transport of chemical elements. We note that the initial values represent both the surface and internal ones, because we assume that stars are born with homogeneous abundance profiles. Moreover, [M/H]$_\mathrm{ini}$=[Fe/H]$_\mathrm{ini}$, and this is valid for every element heavier than helium, because the models include a solar initial metal mixture.  

\section{Optimisation method}
\subsection{AIMS code}

The \textsc{aims}\footnote{\url{https://lesia.obspm.fr/perso/daniel-reese/spaceinn/aims/}} optimisation code (Asteroseismic Inference on a Massive Scale, \citealt{lund18}, \citealt{rendle19}) applies a Markov chain Monte Carlo (MCMC) approach in order to find a representative sample of stellar models that fit a given set of classic and seismic constraints. This sample is subsequently used to find the best-fitting values, error bars, and posterior probability distribution functions (PDFs) for the different stellar properties. In order to gain computation time, the \textsc{aims} code uses a precomputed grid of stellar models which includes global stellar properties such as mass and age, as well as pulsation spectra, and interpolates within this grid for each MCMC iteration. Interpolation is carried out using a multi-dimensional Delaunay tessellation between evolutionary tracks and a simple linear interpolation along evolutionary tracks. \textsc{aims} allows the inclusion of the large frequency separation as a constraint and computes the model value, $\Delta\nu_{\rm mod}$, using the radial modes available in the grid. The posterior distribution of stellar fundamental properties $A$ taking into account observational constraints $O$ is defined as (Bayes' theorem)
\begin{equation}
    p(A|O) \propto p(O|A)p(A),
\end{equation}
\noindent with the likelihood function  
\begin{equation}
    p(O|A) = \frac{1}{\sqrt{2\pi|C|}}\exp(-\chi_\mathrm{tot}^2/2),
\end{equation}
\noindent where $p(A)$ are prior assumptions and $C$ is the covariance matrix of the observed parameters. In this study we assume uniform priors for the stellar fundamental properties. When the large frequency separation is used as a constraint, the $\chi^2$ has the following form:
\begin{equation}
    \chi_\mathrm{tot}^2=\sum\chi_\mathrm{classical}^2+\chi_\mathrm{\Delta\nu}^2,
\end{equation}
\noindent with
\begin{equation}
    \chi^2_i=\left(\frac{O_i-\theta_i}{\sigma_i}\right)^2,
\end{equation}
\noindent where classical means non-seismic constraints ($T_\mathrm{eff}$, luminosity, etc). $O_i$, $\theta_i$, and $\sigma_i$ are respectively the observed value, the model value, and the observational uncertainty. Otherwise the likelihood is only the sum of the classical constraint contributions. 

\subsection{Grid parameters}\label{grid}

\begin{table}
\centering
\caption{Properties of the grid of stellar models used to infer the stellar parameters.} 
\label{table:2}
\begin{tabular}{lcc}
\noalign{\smallskip}\hline\hline\noalign{\smallskip}
Variables  & Range & Steps \\
\noalign{\smallskip}\hline\noalign{\smallskip}
M (M$_\odot$) & $1.2$ - $3.5$ & $0.025$ \\
$Y_\mathrm{ini}$ & $0.242$ - $0.292$ & $0.01$ \\
$Z_\mathrm{ini}$ & $0.0031$ - $0.0381$ & $0.005$ \\
Core overshoot ($H_p$) & $0.0$ - $0.2$ & $0.05$ \\
\noalign{\smallskip}\hline\noalign{\smallskip}
\end{tabular}
\end{table}

The grid of models is Cartesian and includes five dimensions, namely the age, the mass, the initial helium and metal content, and the amount of core overshoot. The details about the grid can be found in Table \ref{table:2}. It has been designed to characterise main sequence stars (minimum hydrogen mass fraction of hydrogen in the core $X_{c}=10^{-11}$). The value of the mixing length parameter is fixed to the solar calibrated value $\alpha_\mathrm{CGM}=0.634$. The initial helium and metal mass fractions are chosen to include the solar calibrated ones as models of the grid ($Y_\mathrm{ini, \odot}=0.252$, and $Z_\mathrm{ini, \odot}=0.0131$). The minimum helium is chosen to include the primordial value and goes slightly below to allow proper probability distributions. The initial metal mass fraction ($Z_\mathrm{ini}$) range is chosen to obtain an initial iron abundance ([Fe/H]$_\mathrm{ini}$) coverage between $-0.5$ and $0.5$~dex. We chose to not include lower metallicities because these are not expected for Ap main-sequence stars. Both $Y_\mathrm{ini}$ and $Z_\mathrm{ini}$ vary freely, that is no enrichment law is taken into account, to prevent any bias from such an assumption. The counter part is the large number of models to compute. The total number of evolutionary tracks in our grid is 21689, including about 6.6 million models (between 200 and 1200 per track, depending on the mass and chemical composition). Core overshoot is chosen to cover the typical values for this parameter in the considered range of mass \citep{claret19}. Oscillation frequencies are computed for each model of the grid using ADIPLS \citep{christensen08} applying a fully reflective boundary condition ($\delta P=0$) at the top of the atmosphere. This boundary condition ensures that the frequencies are computed up to the observed values, which are greater than the acoustic cutoff frequency in both stars for which seismic constraints are applied. These very high frequencies are observed in roAp stars because the magnetic field provides a mechanism for partial refraction of the modes at frequencies higher than the acoustic cutoff frequency \citep{sousa08,quitral18}.

\subsection{Classical constraints}

We used the radius and the luminosity as classical constraints. We do not consider the effective temperature, because the latter is directly related  to the other two through the Stefan-Boltzmann law and does not add an independent constraint. A wide variety of probability distribution functions can be applied by \textsc{aims}. In what follows, we used normal distributions, truncated at 3$\sigma$, for the two classical constraints.

As discussed in Section~\ref{models}, the stellar models we use to determine the stellar parameters include no transport of chemical element. As [Fe/H] can be strongly affected by magnetic fields and transport processes \citep[e.g.][]{shulyak09}, we cannot use it as classical constraint.

\subsection{Seismic constraints}\label{sismo}

We use the large frequency separations observed in HR\,1217 (HD\,24712) and $\alpha$~Cir (HD\,128898) to improve the fits, and assess the effect this additional constraint has on the probability distributions of the inferred stellar fundamental properties. Following on the discussion in Section~\ref{seismology}, we consider two different scenarios. First we employ the constraint $\Delta\nu_{\rm mag}$, which incorporates a correction due the possible impact of the magnetic field. Under this scenario, we further consider two possibilities for the probability distribution of $\Delta\nu_{\rm mag}$, namely, a normal distribution with a central value of $(\Delta\nu-1.0)$ and standard deviation of $2.0~\mu\mathrm{Hz}$ (hereafter $\Delta\nu(a)$) and a uniform distribution with a range [$\Delta\nu-3.0$ ; $\Delta\nu+1.0$]~$\mu\mathrm{Hz}$ (hereafter $\Delta\nu(b)$). The second scenario disregards the potential effect of the magnetic field, considering a normal distribution for the observed values $\Delta\nu_{\rm obs}$ (hereafter $\Delta\nu(c)$).  

\section{Parameter inferences}

\subsection{Masses, $X_c$, and ages from classical constraints}

\begin{figure*}[!ht]
\center
\includegraphics[scale=0.82]{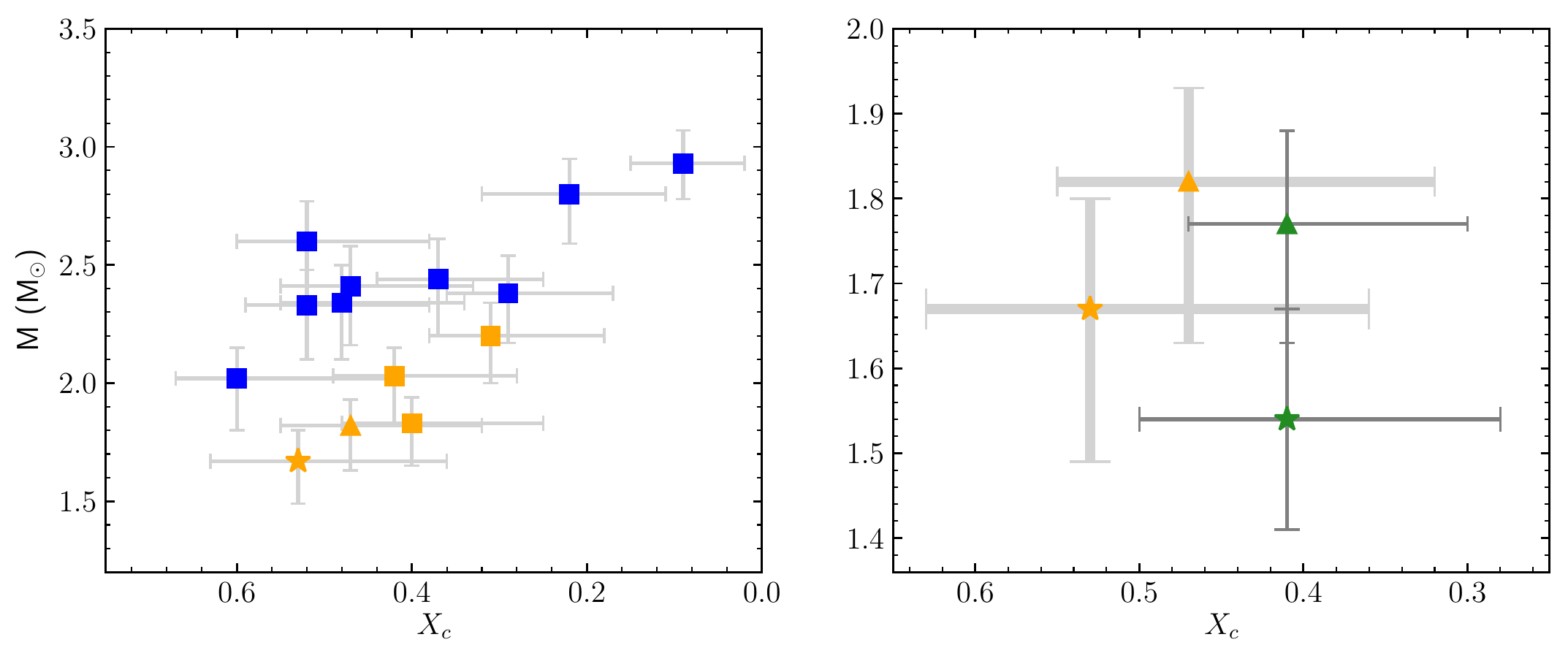}
\caption{\textit{Left:} Inferred mass according to the inferred hydrogen mass fraction in the core ($X_c$) using the classical constraints, for the 14 stars of the sample. The orange symbols represent the roAp stars of the sample while the blue ones represent the other stars of the sample. The star and pyramid symbols represent respectively HR\,1217 (HD\,24712) and $\alpha$~Cir (HD\,128898). \textit{Right:} Same as the left panel for HR\,1217 (HD\,24712) and $\alpha$~Cir (HD\,128898) (orange symbols). Green symbols are the inferences when the $\Delta\nu(b)$ seismic constrain is taken into account.}
\label{fig:M-Xc}
\end{figure*}

\begin{table*}[ht]
\centering
\caption{Stellar parameters obtained with \textsc{aims}. Uncertainties are given at $1\sigma$ ($16^\mathrm{th}$ and $84^\mathrm{th}$ percentiles).} 
\label{table:3}
\begin{tabular}{l|ccc|ccc}
\noalign{\smallskip}\hline\hline\noalign{\smallskip}
& \multicolumn{3}{c|}{Free d$Y$/d$Z$} & \multicolumn{3}{c}{d$Y$/d$Z$$~=[0.4,3.0]$} \\
\noalign{\smallskip}\hline\noalign{\smallskip}
HD  & Mass [M$_\odot$] & $X_c$ & Age [Gyr] & Mass [M$_\odot$] & $X_c$ & Age [Gyr]\\
\noalign{\smallskip}\hline\noalign{\smallskip}
4778   & $2.41^{+0.17}_{-0.25}$ & $0.45^{+0.09}_{-0.14}$ & $0.38^{+0.19}_{-0.13}$ & $2.39^{+0.15}_{-0.19}$ & $0.45^{+0.08}_{-0.11}$ & $0.38^{+0.16}_{-0.13}$\\
%4778   & $2.42^{+0.17}_{-0.24}$ & $0.47^{+0.08}_{-0.14}$ & $0.37^{+0.18}_{-0.12}$ & $2.39^{+0.15}_{-0.19}$ & $0.45^{+0.07}_{-0.11}$ & $0.38^{+0.16}_{-0.12}$\\
\noalign{\smallskip}
\noalign{\smallskip}\hline\noalign{\smallskip}
24712  & $1.67^{+0.13}_{-0.18}$ & $0.53^{+0.10}_{-0.17}$ & $0.78^{+0.76}_{-0.47}$ & $1.66^{+0.13}_{-0.16}$ & $0.52^{+0.09}_{-0.15}$ & $0.80^{+0.68}_{-0.45}$ \\
\noalign{\smallskip}
$\Delta\nu(a)$  & $1.55^{+0.13}_{-0.14}$ & $0.41^{+0.10}_{-0.15}$ & $1.29^{+0.60}_{-0.44}$  & $1.56^{+0.12}_{-0.12}$ & $0.42^{+0.09}_{-0.13}$ & $1.20^{+0.59}_{-0.39}$\\
\noalign{\smallskip}
$\Delta\nu(b)$  & $1.54^{+0.13}_{-0.13}$ & $0.39^{+0.09}_{-0.13}$ & $1.32^{+0.57}_{-0.40}$ & $1.55^{+0.12}_{-0.12}$ & $0.40^{+0.08}_{-0.11}$ & $1.27^{+0.54}_{-0.36}$ \\
\noalign{\smallskip}
$\Delta\nu(c)$  & $1.58^{+0.11}_{-0.12}$ & $0.43^{+0.08}_{-0.10}$ & $1.19^{+0.35}_{-0.31}$ & $1.57^{+0.10}_{-0.09}$ & $0.43^{+0.07}_{-0.08}$ & $1.17^{+0.33}_{-0.30}$ \\
\noalign{\smallskip}
\noalign{\smallskip}\hline\noalign{\smallskip}
\noalign{\smallskip}
108662 & $2.44^{+0.17}_{-0.24}$ & $0.38^{+0.09}_{-0.13}$ & $0.45^{+0.15}_{-0.11}$ & $2.40^{+0.16}_{-0.19}$ & $0.36^{+0.08}_{-0.11}$ & $0.46^{+0.13}_{-0.11}$\\
%108662 & $2.40^{+0.16}_{-0.22}$ & $0.37^{+0.07}_{-0.12}$ & $0.48^{+0.15}_{-0.10}$ & $2.38^{+0.15}_{-0.18}$ & $0.36^{+0.07}_{-0.10}$ & $0.48^{+0.13}_{-0.10}$\\
\noalign{\smallskip}
108945 & $2.38^{+0.16}_{-0.21}$ & $0.29^{+0.08}_{-0.12}$ & $0.56^{+0.13}_{-0.10}$ & $2.37^{+0.13}_{-0.18}$ & $0.28^{+0.07}_{-0.10}$ & $0.56^{+0.12}_{-0.09}$\\
%108945 & $2.38^{+0.15}_{-0.21}$ & $0.29^{+0.07}_{-0.12}$ & $0.57^{+0.12}_{-0.10}$ & $2.36^{+0.14}_{-0.19}$ & $0.28^{+0.06}_{-0.11}$ & $0.56^{+0.13}_{-0.09}$\\
\noalign{\smallskip}
118022 & $2.33^{+0.15}_{-0.23}$ & $0.52^{+0.07}_{-0.14}$ & $0.38^{+0.22}_{-0.13}$ & $2.30^{+0.14}_{-0.18}$ & $0.50^{+0.07}_{-0.10}$ & $0.34^{+0.18}_{-0.13}$\\
\noalign{\smallskip}
120198 & $2.60^{+0.17}_{-0.25}$ & $0.53^{+0.08}_{-0.13}$ & $0.24^{+0.15}_{-0.11}$ & $2.57^{+0.16}_{-0.21}$ & $0.52^{+0.08}_{-0.11}$ & $0.24^{+0.13}_{-0.10}$\\
%120198 & $2.60^{+0.17}_{-0.26}$ & $0.52^{+0.08}_{-0.14}$ & $0.24^{+0.16}_{-0.10}$ & $2.58^{+0.15}_{-0.20}$ & $0.51^{+0.07}_{-0.11}$ & $0.24^{+0.12}_{-0.09}$\\
\noalign{\smallskip}
\noalign{\smallskip}\hline\noalign{\smallskip}
128898 & $1.82^{+0.11}_{-0.19}$ & $0.47^{+0.08}_{-0.15}$ & $0.81^{+0.43}_{-0.25}$ & $1.80^{+0.10}_{-0.14}$ & $0.45^{+0.08}_{-0.11}$ & $0.82^{+0.35}_{-0.24}$\\
\noalign{\smallskip}
$\Delta\nu(a)$ & $1.77^{+0.12}_{-0.15}$ & $0.42^{+0.07}_{-0.12}$ &
$0.97^{+0.30}_{-0.23}$ & $1.76^{+0.11}_{-0.13}$ & $0.41^{+0.06}_{-0.10}$ & $0.95^{+0.27}_{-0.21}$\\
\noalign{\smallskip}
$\Delta\nu(b)$ & $1.77^{+0.11}_{-0.14}$ & $0.41^{+0.06}_{-0.11}$ &
$0.98^{+0.27}_{-0.19}$ & $1.75^{+0.11}_{-0.12}$ & $0.40^{+0.06}_{-0.09}$ & $0.97^{+0.24}_{-0.20}$\\
\noalign{\smallskip}
$\Delta\nu(c)$ & $1.80^{+0.09}_{-0.13}$ & $0.44^{+0.05}_{-0.08}$ &
$0.91^{+0.23}_{-0.16}$ & $1.77^{+0.10}_{-0.11}$ & $0.43^{+0.05}_{-0.07}$ & $0.90^{+0.22}_{-0.16}$ \\
\noalign{\smallskip}
\noalign{\smallskip}\hline\noalign{\smallskip}
137909 & $2.20^{+0.14}_{-0.20}$ & $0.31^{+0.07}_{-0.13}$ & $0.70^{+0.17}_{-0.13}$ & $2.17^{+0.13}_{-0.17}$ & $0.30^{+0.07}_{-0.11}$ & $0.70^{+0.16}_{-0.12}$\\
\noalign{\smallskip}
153882 & $2.80^{+0.15}_{-0.21}$ & $0.22^{+0.10}_{-0.11}$ & $0.41^{+0.07}_{-0.06}$ & $2.77^{+0.14}_{-0.20}$ & $0.22^{+0.12}_{-0.12}$ & $0.39^{+0.08}_{-0.08}$\\
\noalign{\smallskip}
176232 & $2.03^{+0.12}_{-0.20}$ & $0.42^{+0.07}_{-0.14}$ & $0.69^{+0.27}_{-0.17}$ & $2.01^{+0.11}_{-0.17}$ & $0.41^{+0.07}_{-0.11}$ & $0.69^{+0.24}_{-0.17}$\\
\noalign{\smallskip}
188041 & $2.34^{+0.16}_{-0.24}$ & $0.48^{+0.07}_{-0.14}$ & $0.39^{+0.21}_{-0.13}$ & $2.32^{+0.15}_{-0.18}$ & $0.47^{+0.06}_{-0.10}$ & $0.38^{+0.16}_{-0.12}$\\
\noalign{\smallskip}
201601 & $1.83^{+0.11}_{-0.18}$ & $0.40^{+0.08}_{-0.15}$ & $0.99^{+0.36}_{-0.24}$ & $1.80^{+0.10}_{-0.16}$ & $0.38^{+0.08}_{-0.12}$ & $1.00^{+0.33}_{-0.24}$\\
\noalign{\smallskip}
204411 & $2.93^{+0.14}_{-0.15}$ & $0.10^{+0.05}_{-0.07}$ & $0.43^{+0.06}_{-0.05}$ & $2.89^{+0.14}_{-0.15}$ & $0.09^{+0.05}_{-0.07}$ & $0.44^{+0.07}_{-0.05}$\\
%204411 & $2.91^{+0.14}_{-0.16}$ & $0.09^{+0.06}_{-0.07}$ & $0.43^{+0.06}_{-0.05}$ & $2.89^{+0.13}_{-0.15}$ & $0.08^{+0.05}_{-0.06}$ & $0.43^{+0.06}_{-0.05}$\\
\noalign{\smallskip}
220825 & $2.02^{+0.13}_{-0.22}$ & $0.60^{+0.07}_{-0.17}$ & $0.28^{+0.38}_{-0.20}$ & $2.04^{+0.11}_{-0.14}$ & $0.61^{+0.06}_{-0.10}$ & $0.24^{+0.24}_{-0.16}$\\
\noalign{\smallskip}\hline\noalign{\smallskip}
\end{tabular}
\end{table*}

In this section we discuss the results of the property inferences with \textsc{aims}. Table \ref{table:3} gives the masses, hydrogen mass fractions in the core ($X_c$), and the ages of the 14 stars when $Y$ and $Z$ vary freely (Free d\textit{Y}/d\textit{Z}, first column), and when $Y$ varies according to $Z$ following an enrichment law (second column). We first discuss the results with a free d\textit{Y}/d\textit{Z}, the second case is detailed in Section~\ref{enrich}. The corresponding probability density functions are presented in Appendix \ref{Appen}, and in Fig.~\ref{fig:HD24712} and \ref{fig:HD128898} for HR\,1217 (HD\,24712) and $\alpha$~Cir (HD\,128898) respectively. The probability density functions are not strictly normal and present an asymmetry (positive or negative depending on the parameter). The central values given in Table \ref{table:3} are the medians of the distributions (i.e. the 50$^\mathrm{th}$ percentile). We define the asymmetric uncertainties at $1\sigma$ with the 16$^\mathrm{th}$ and 84$^\mathrm{th}$ percentile of the distributions.

Figure \ref{fig:M-Xc} (left panel) shows the mass and hydrogen mass fraction in the core ($X_c$) inferred for the stars of the sample. Taking into account the $1\sigma$ error bars, the stars have masses between $1.5$ and $3.0$~M$_\odot$. Most of the stars are in the first half of the main sequence with $X_c>0.35$ and ages between about $0.1$ and $1.6$~Gyr, except for HD\,204411 which seem to be close to the end of main sequence with $X_c=0.10^{+0.05}_{-0.07}$. The uncertainties on $X_c$ for HD\,204411 are smaller than for the others. Its relative errors is less important because it is possibly a subgiant and the minimum hydrogen mass fraction of the models in the grid is $X_c=10^{-11}$, which in the middle of the subgiant phase. A more accurate inference of the stellar fundamental properties of HD\,204411 would require extending the grid into the subgiant phase.

The inferred ages should be interpreted as estimates only. The grid of models includes core overshoot, but neglects other processes that may significantly affect the lifetime of main-sequence stars. Studies of G and F-type stars using grid-based modelling and including acoustic seismic constraints show that atomic diffusion can impact the age of a star by up to $15\%$ \citep[][]{nsamba18,deal20}. Similar results were also found for A-type stars using gravity modes as constraints \citep{mombarg20}. The effect of rotation is also not taken into account in the models. The transport of chemical elements by rotation has a direct impact on the lifetime of a star on the main sequence. There is also a degeneracy between the effect of core overshoot and rotation on the stellar age \cite[e.g.][]{maeder09}. In this context, $X_c$ is more reliable than age as it is a measure of the fraction of evolution on the main sequence.

\subsection{Chemical composition}

Our results show that the chemical composition of the stars under study is not constrained by the classical observations alone. There is a clear degeneracy both in the helium and metal contents. In the case of helium, we find that the probability density functions are homogeneous for all $Y_\mathrm{ini}$ values of the grid (see lower left panel of Fig. \ref{fig:HD24712} for the example of HR\,1217/HD\,24712). For [M/H]$_\mathrm{ini}$, the probability density functions are pointing towards the higher metallicities of the grid (see Appendix \ref{Appen}) with no metallicities excluded (except, for HD\,204411 for the reasons previously mentioned). This shape of the distribution of [M/H]$_\mathrm{ini}$, is expected for a Cartesian grid in $Z_\mathrm{ini}$ (see Appendix \ref{distribFeH}). It means that [Fe/H]$_\mathrm{ini}$ is not constrained at all by the radius and luminosity.

The $Z_\mathrm{ini}$ probability density functions are not completely homogeneous and show an increasing trend towards high initial metal mass fraction (less pronounced than [M/H]$_\mathrm{ini}$) as shown on the lower middle panel of Fig \ref{fig:HD24712} for HR\,1217 (HD\,24712). Contrarily to [M/H]$_\mathrm{ini}$, this is not expected. This comes from the architecture of the grid (all models are stopped at $X_c=10^{-11}$), which was designed to look for main sequence stars. At low metallicities, we can see on the bottom panel of Fig. \ref{fig:HR} that the stars are at the edge of the tracks (outside for HD\,153882 and HD\,204411). This is the case at every $Y_\mathrm{ini}$. It means that there are less potential fitting models at lower than at higher metallicities, where all stars lie well within the evolutionary tracks. We note that we do not expect these young stars to have such low metallicities. We can thus consider $Z_\mathrm{ini}$, similarly to [M/H]$_\mathrm{ini}$, completely unconstrained, taking into account the main sequence assumption.

\subsection{Benefits from seismic constraints}\label{result-sismo}

\begin{figure*}[!ht]
\center
\includegraphics[scale=0.75]{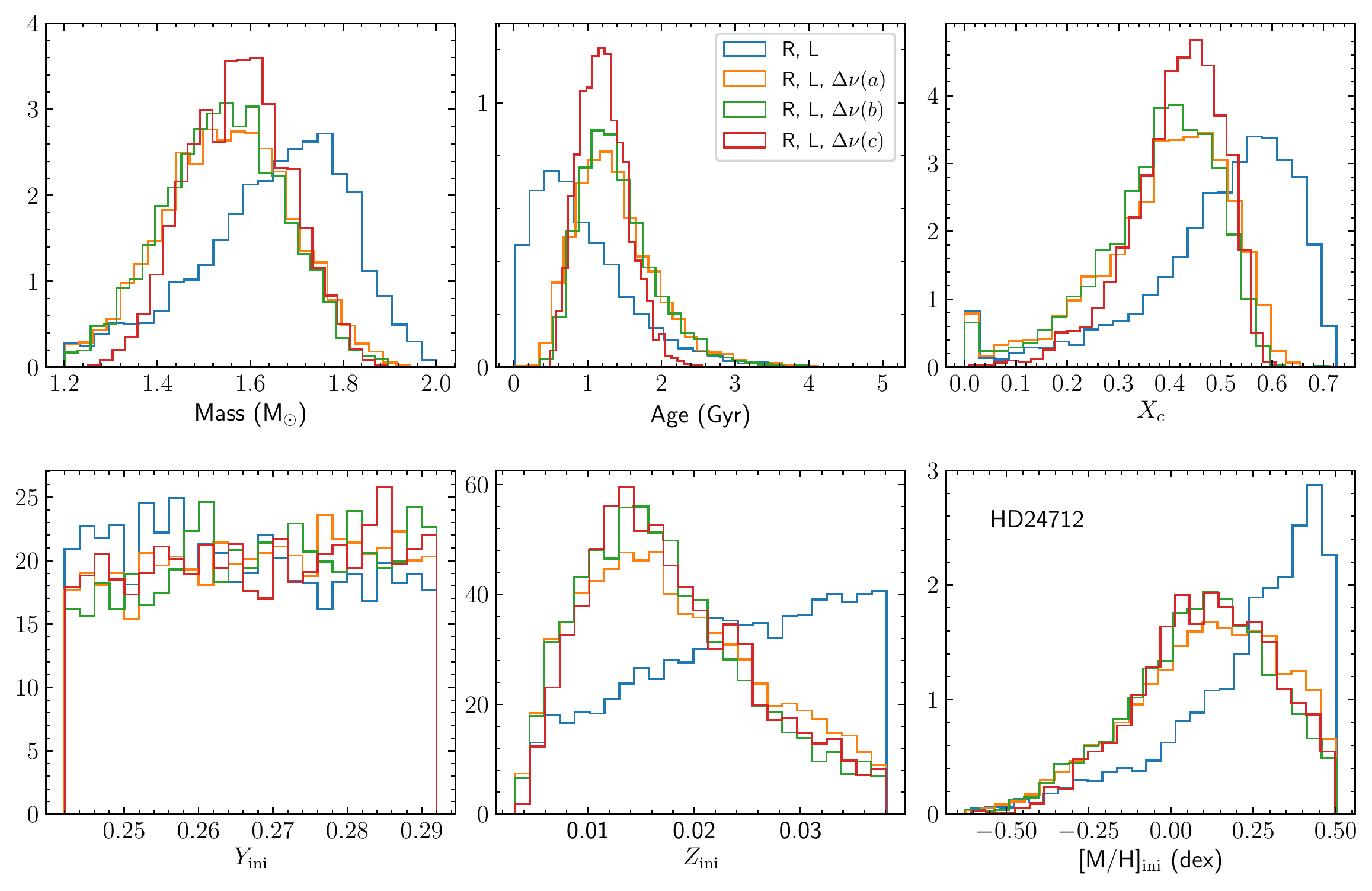}
\caption{Probability density functions for the mass, age, $X_c$, $Y_\mathrm{ini}$, $Z_\mathrm{ini}$, and [M/H]$_\mathrm{ini}$ of HR\,1217 (HD\,24712). The blue distributions take into account the classical constraints only, while the orange, green, and red ones include $\Delta\nu$ as an additional constraint with different uncertainties (see Section \ref{sismo}).}
\label{fig:HD24712}
\end{figure*}

\begin{figure*}[ht]
\center
\includegraphics[scale=0.75]{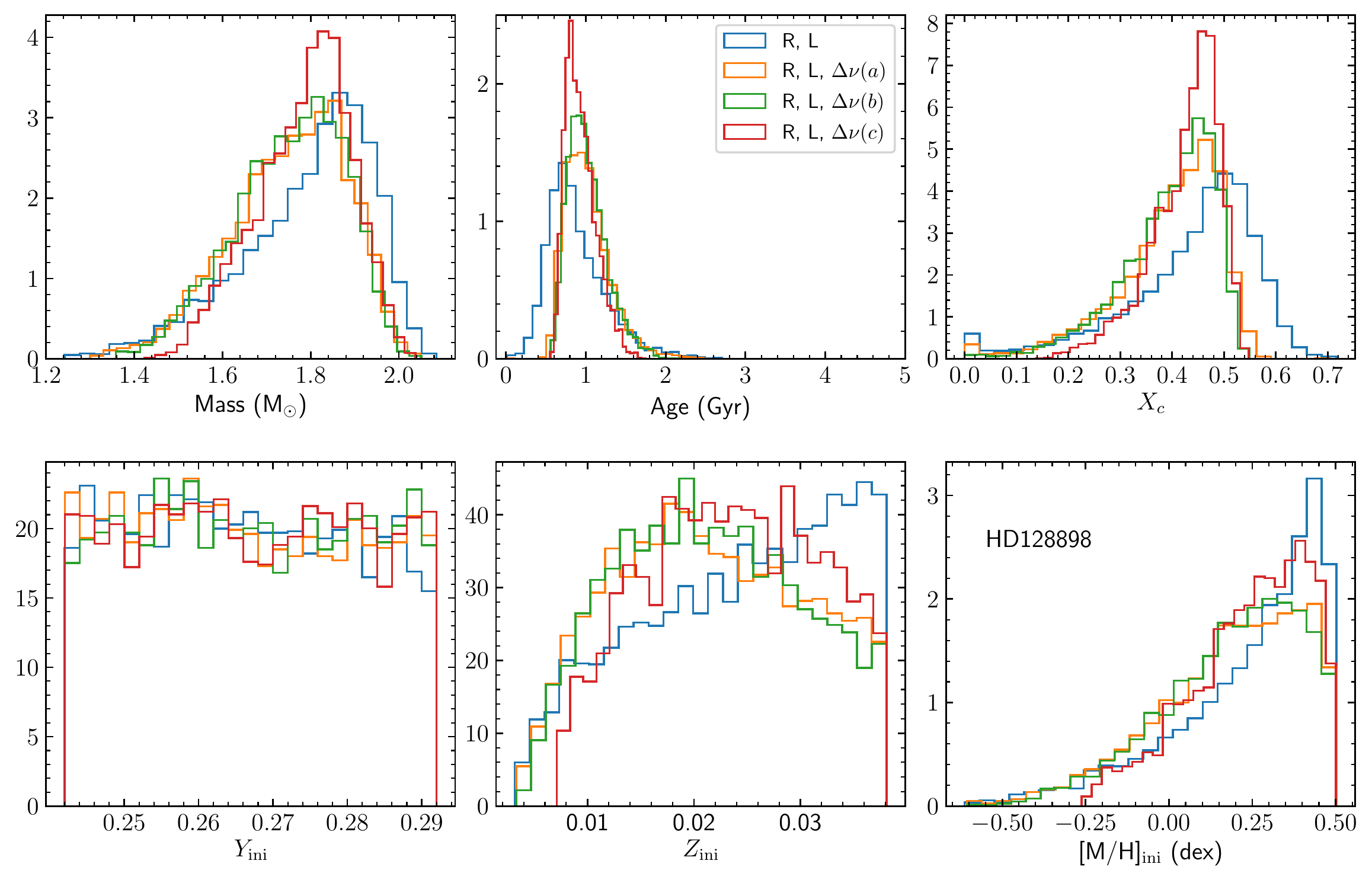}
\caption{Same as Fig.~\ref{fig:HD24712}, but for $\alpha$~Cir (HD\,128898).}
\label{fig:HD128898}
\end{figure*}

For two stars of the sample (HR\,1217/HD\,24712 and $\alpha$~Cir/HD\,128898) we have a constraint on $\Delta\nu$, in addition to the radius and luminosity. Figures \ref{fig:HD24712} and \ref{fig:HD128898} show the impact on the property inferences of adding this seismic constraint with uncertainties as described in Section \ref{sismo}, for HR\,1217 (HD\,24712) and $\alpha$~Cir (HD\,128898) respectively. We can see that all three stellar properties are better constrained when the seismic information is considered, regardless of the type of uncertainties we use on $\Delta\nu_{\rm mag}$ (Gaussian or uniform, cases a and b, respectively). As expected, the constraining power of the seismic information increases when the more precise observed (uncorrected) value $\Delta\nu_{\rm obs}$, with smaller uncertainties, is adopted (case c). The values of the inferred masses, ages, and $X_c$ for the different cases considered for HR\,1217 (HD\,24712) and $\alpha$~Cir (HD\,128898) are listed in Table \ref{table:3}. As expected, the main improvement brought by the seismic constraint is on the metal content. When $\Delta\nu$ is taken into account, the distribution in $Z_\mathrm{ini}$ (and in [M/H]$_\mathrm{ini}$) is better constrained. When uniform conservative uncertainties (cases b) are considered, we obtain $1\sigma$ intervals of $Z_\mathrm{ini}=0.016^{+0.009}_{-0.007}$ and [M/H]$_\mathrm{ini}~=0.10^{+0.20}_{-0.23}$~dex for HR\,1217 (HD\,24712), and $Z_\mathrm{ini}=0.021^{+0.010}_{-0.009}$ and [M/H]$_\mathrm{ini}~=0.22^{+0.18}_{-0.25}$~dex for $\alpha$~Cir (HD\,128898). For all stellar properties, the improvement from adding the seismic constraint is more significant for HR\,1217 (HD\,24712) than for $\alpha$~Cir (HD\,128898). This is due to a smaller uncertainty on the luminosity for $\alpha$~Cir (HD\,128898), inducing a stronger constraint on the stellar fundamental properties without seismic constraint.\newline

\textbf{Abundances of Si, Ca, and Fe of HR\,1217 (HD\,24712):}

Abundance stratification of several elements have been determined for HR\,1217 (HD\,24712) by \cite{shulyak09}. The elements in common with the ones followed in \textsc{CESTAM} models are Si, Ca, and Fe. In the lowest atmospheric layers they were able to probe ($\log_{10}\tau_{5000}\simeq0$), they obtained $\log_{10}(N_\mathrm{Si}/N_\mathrm{tot})=-3.60$, $\log_{10}(N_\mathrm{Ca}/N_\mathrm{tot})=-4.60$, and $\log_{10}(N_\mathrm{Fe}/N_\mathrm{tot})=-4.25$. Using the solar abundance of \cite{asplund09} (ie. $\log_{10}(N_\mathrm{Si}/N_\mathrm{tot})_\odot=-4.53$, $\log_{10}(N_\mathrm{Ca}/N_\mathrm{tot})_\odot=-5.70$, and $\log_{10}(N_\mathrm{Fe}/N_\mathrm{tot})_\odot=-4.54$), we obtain [Si/H]$_\mathrm{HD\,24712}=0.93$~dex, [Ca/H]$_\mathrm{HD\,24712}=1.10$~dex, and [Fe/H]$_\mathrm{HD\,24712}=0.29$~dex. Our models predict [M/H]$_\mathrm{ini}$=[Si/H]$_\mathrm{ini}$=[Ca/H]$_\mathrm{ini}$=[Fe/H]$_\mathrm{ini}=0.10^{+0.20}_{-0.23}$~dex.

In all three cases, the observed abundances at $\log_{10}\tau_{5000}\simeq0$ are larger than the predicted initial abundances (in the case of Si and Ca, the difference being rather significant). This could be an indication of the effect of atomic diffusion in the interior of Ap stars. 

As a test of the impact of the processes leading to chemical transport, we computed a model with the same evolution code (\textsc{cestam}), including the effect of atomic diffusion (with radiative acceleration) as described in \cite{deal18}, and an additional parametrised transport process similar to what is expected in Am stars \citep[e.g.][]{richer00,richard01,michaud11}. A more detailed description of the physics of this model is presented in Section~\ref{discu-diff}. We used the median values obtained with the constrain of $\Delta\nu(b)$ and a free d$Y$/d$Z$ to compute the model (M=1.54~M$_\odot$, $X_c=0.39$ and [Fe/H]$_\mathrm{ini}=0.10$~dex). The models predict [Si/H]=-0.13~dex, [Ca/H]=-0.63~dex and [Fe/H]=0.58~dex. Si and Ca surface abundances are lower than the initial ones while iron surface abundance is larger. This is what we expect from models including atomic diffusion in the stellar interior \citep[e.g.][]{deal18}. These predicted abundances at the bottom of the atmosphere give an indication of the reservoir of chemical element available in the atmosphere. The fact that the predicted iron abundance at the bottom of the atmosphere is 0.5~dex larger than the initial one indicates that the available quantity of iron is sufficient to explain the stratification obtained from observations. For the two other elements, it indicates that the macroscopic transport of chemical elements may be more efficient in Ap stars than in Am stars, in order to prevent the depletion of these element from the surface.\newline

\textbf{Abundances of Si, Ca, and Fe of $\alpha$ Cir (HD\,128898):}

We performed a similar analysis for $\alpha$ Cir (HD\,128898) for which we determined [M/H]$_\mathrm{ini}$=[Si/H]$_\mathrm{ini}$=[Ca/H]$_\mathrm{ini}$=[Fe/H]$_\mathrm{ini}=0.22^{+0.18}_{-0.25}$~dex. We use the abundance stratification determined by \cite{kochukhov09} and obtained (at $\log_{10}\tau_{5000}\simeq0$) [Si/H]$_\mathrm{HD\,128898}=1.03$~dex, [Ca/H]$_\mathrm{HD\,128898}=1.80$~dex, and [Fe/H]$_\mathrm{HD\,128898}=0.69$~dex.

Similarly to $\alpha$ Cir (HD\,128898), we computed a model including the effect of atomic diffusion (with radiative acceleration) and an additional parametrised transport process similar to what is expected in Am stars. We used the median values obtained with the constrain of $\Delta\nu(b)$ and a free d$Y$/d$Z$ to compute the model (M=1.77~M$_\odot$, $X_c=0.41$, and [Fe/H]$_\mathrm{ini}=0.22$~dex). We obtained [Si/H]=0.01~dex, [Ca/H]=-0.39~dex, and [Fe/H]=0.69~dex. This leads to the same conclusion as for HR\,1217~(HD\,24712).

The stratification of chemical elements was also derived for other stars of the sample (HD\,176232/10 Aql: \citealt{nesvacil13} ; HD\,137909/$\beta$ CrB and HD\,201601/$\gamma$ Equ A: \citealt{shulyak13} ; HD\,204411: \citealt{ryabchikova05}). Because [M/H]$_\mathrm{ini}$ is not well constrained for these stars, we cannot perform a similar comparison.\newline

\textbf{Mass range of Ap stars:}

When only classical constraints are taken into account, the mass probability density functions of HR\,1217 (HD\,24712) and $\alpha$~Cir (HD\,128898) indicate at $3\sigma$ a minimum mass of $1.21$ and $1.28$~M$_\odot$, respectively. It changes to $1.21$ and $1.38$~M$_\odot$ when $\Delta\nu$ with uniform uncertainty is considered. These masses are smaller than the current mass range typically assumed for Ap stars, which often starts in 1.5~M$_\odot$, based on the assumption of a solar chemical composition. The fact that the initial metallicity is largely unconstrained, leading to a significant uncertainty in the mass, indicates that the minimum mass of Ap stars may need to be shifted to lower values.

\subsection{Enrichment law}\label{enrich}

\begin{figure*}
\center
\includegraphics[scale=0.80]{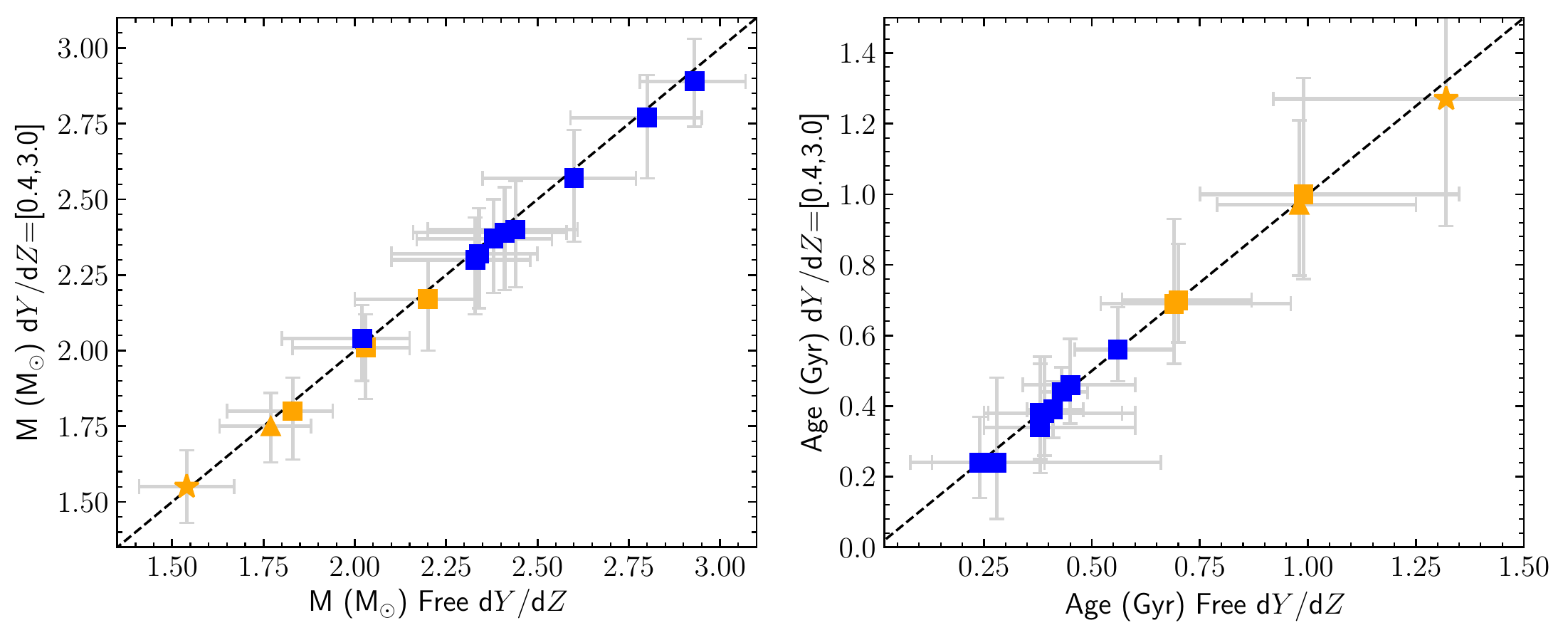}
\caption{Star-by-star comparison of the inferred mass (left panel) and age (right panel) using a free value of $\rm{d}Y/\rm{d}Z$ or an observationaly constrained one. The orange symbols are the five roAp stars of the sample while the blue ones are for the others. The star and pyramid symbols represent respectively the properties for HR\,1217 (HD\,24712) and $\alpha$~Cir (HD\,128898) inferred with the additional constraint $\Delta\nu(b)$. The dashed black lines are the 1:1 comparisons }
\label{fig:dydz}
\end{figure*}

An enrichment law characterises the way the abundance of helium varies with the metallicity. It is suitable to study ensembles of stars but is less relevant for studies of individual stars, especially population I stars where the dispersion in this relation can be significant \citep[e.g.][]{verma19}. This is the reason why we decided to not assume any enrichment law to start with. Nevertheless, we tested the impact of considering a constrained enrichment law. According to observations, the helium-to-heavy element enrichment ratio $\rm{d}Y/\rm{d}Z$ ranges between $0.4$ and $3$ \citep[see][for a review]{nsamba21}. The impact on the posterior distributions from constraining the value of d\textit{Y}/d\textit{Z} can be seen in Table~\ref{table:3} and in Fig.~\ref{fig:dydz}. The median values for mass are slightly smaller when d\textit{Y}/d\textit{Z} is constrained while remaining very close for $X_c$ and age. As an additional constraint is taken into account, the $1\sigma$ uncertainties are smaller, as expected.

\section{Impact from additional physical processes}

As we neglected both chemical element transport processes and the effect of magnetic fields, we assess in this Section the impact they have on the results. We show that in both cases, the impact on the stellar properties is negligible compared to the uncertainties of the observational constraints. In what concerns the impact of chemical element transport, we stress that this is only valid because we did not include surface abundances as an observational constraint in the analysis. When such constraints are used, a non-adequate account for chemical transport can bias the inferences, increasing the impact of gravitational settling, especially in the inference of the age \cite[e.g.][]{nsamba18}. Similarly, atomic diffusion (with radiative acceleration) has been shown to have a significant impact on stellar properties when surface abundances are used as constraints \citep[e.g.][]{deal20}. The use of more realistic models accounting for the transport of chemical elements is the next natural step in the context of studies like this one. Such models would allow us to use observational information about the observed bulk chemical abundances and hopefully reduce the uncertainties on the inferred stellar properties. 

\subsection{Impact of chemical element transport processes}\label{discu-diff}

\begin{figure*}[!ht]
\center
\includegraphics[scale=0.71]{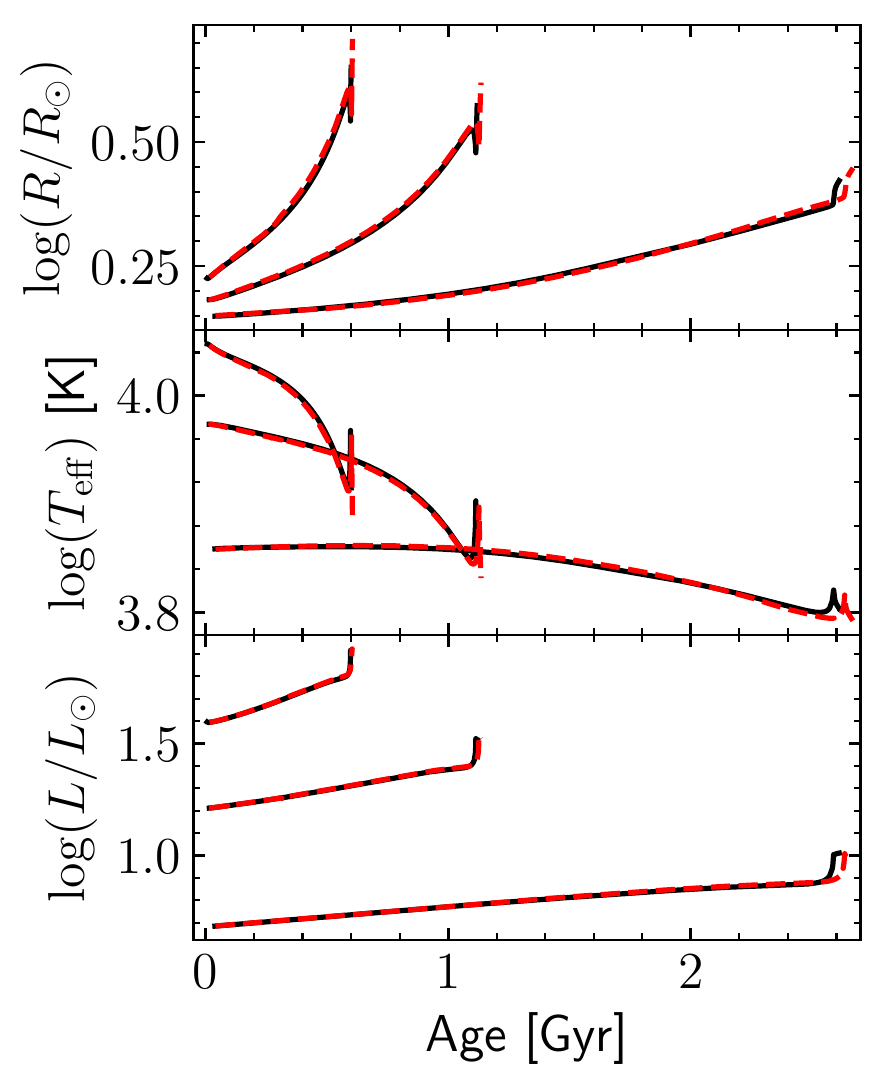}\hspace{1.0cm}\includegraphics[scale=0.7]{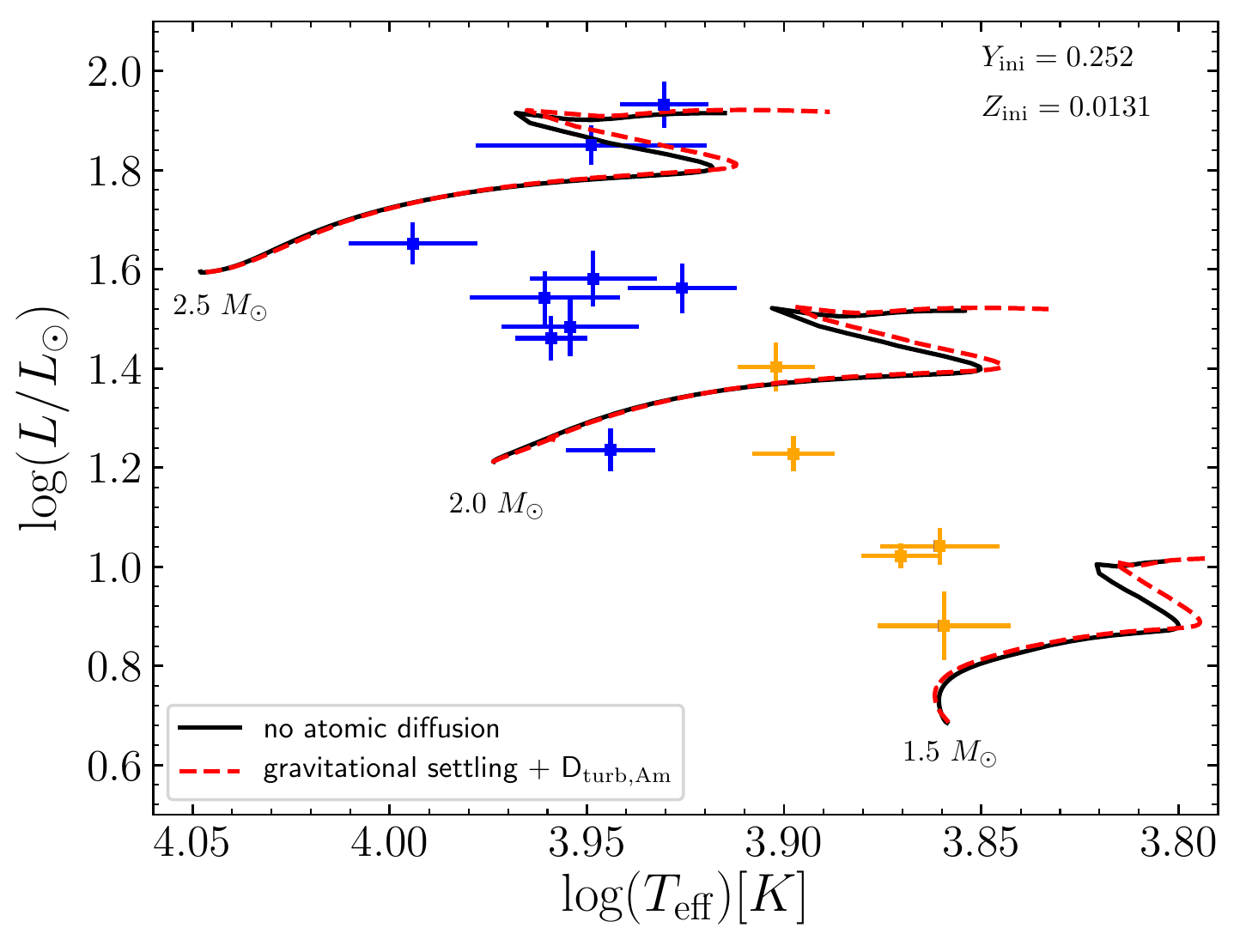}
\caption{Shown are two sets of \textsc{CESTAM} evolutionary models. The first one (black solid lines) with the same physics as the model presented in the middle panel of Fig.~\ref{fig:HR} (representative physics of the grid used in this study). The second one (red dashed lines) including gravitational settling and a parametrised turbulent diffusion coefficient calibrated on Am stars (see text). The blue and orange crosses are the same as in Fig.~\ref{fig:HR}.}
\label{fig:diff}
\end{figure*}

\begin{figure}[!ht]
\center
\includegraphics[scale=0.58]{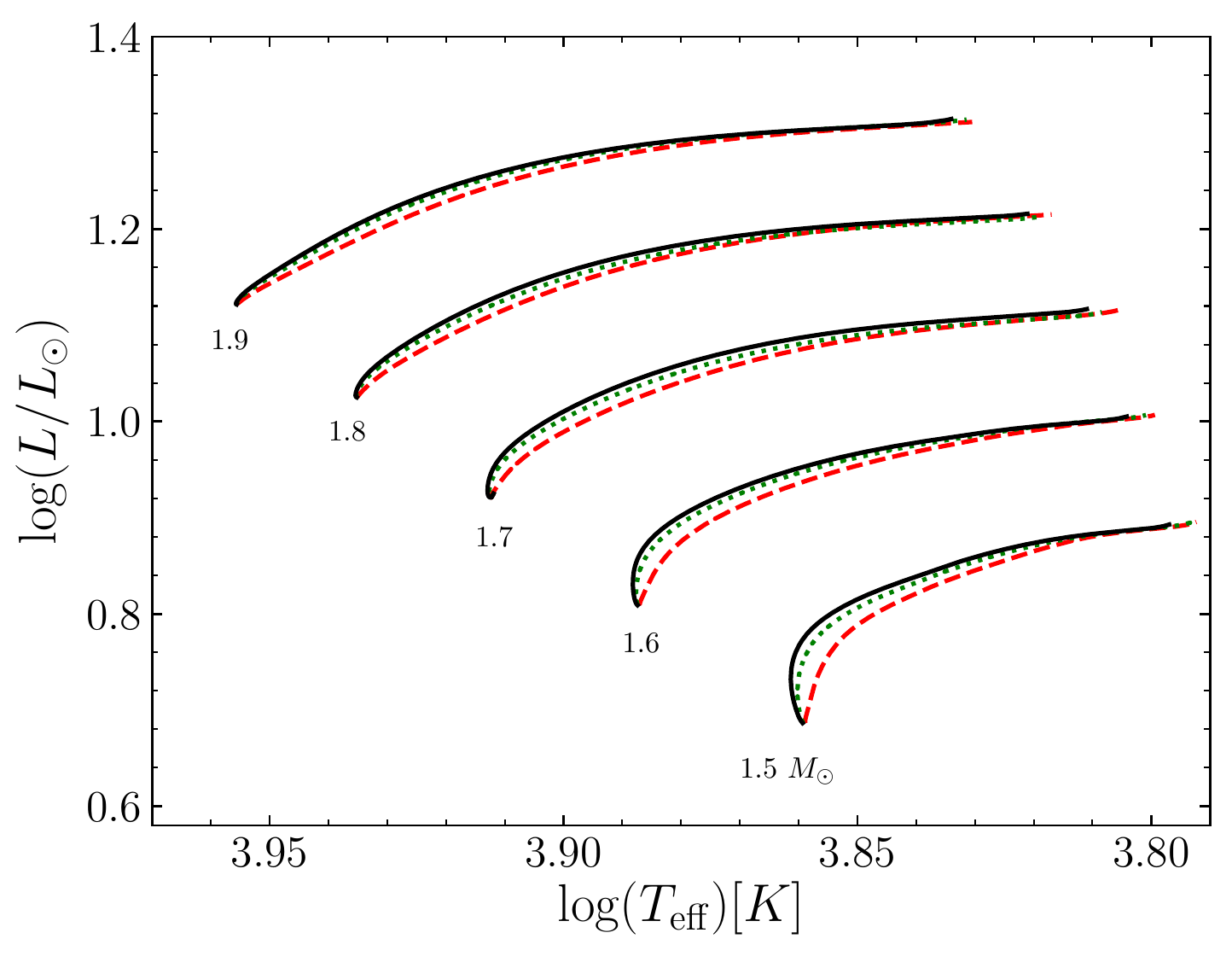}
\includegraphics[scale=0.58]{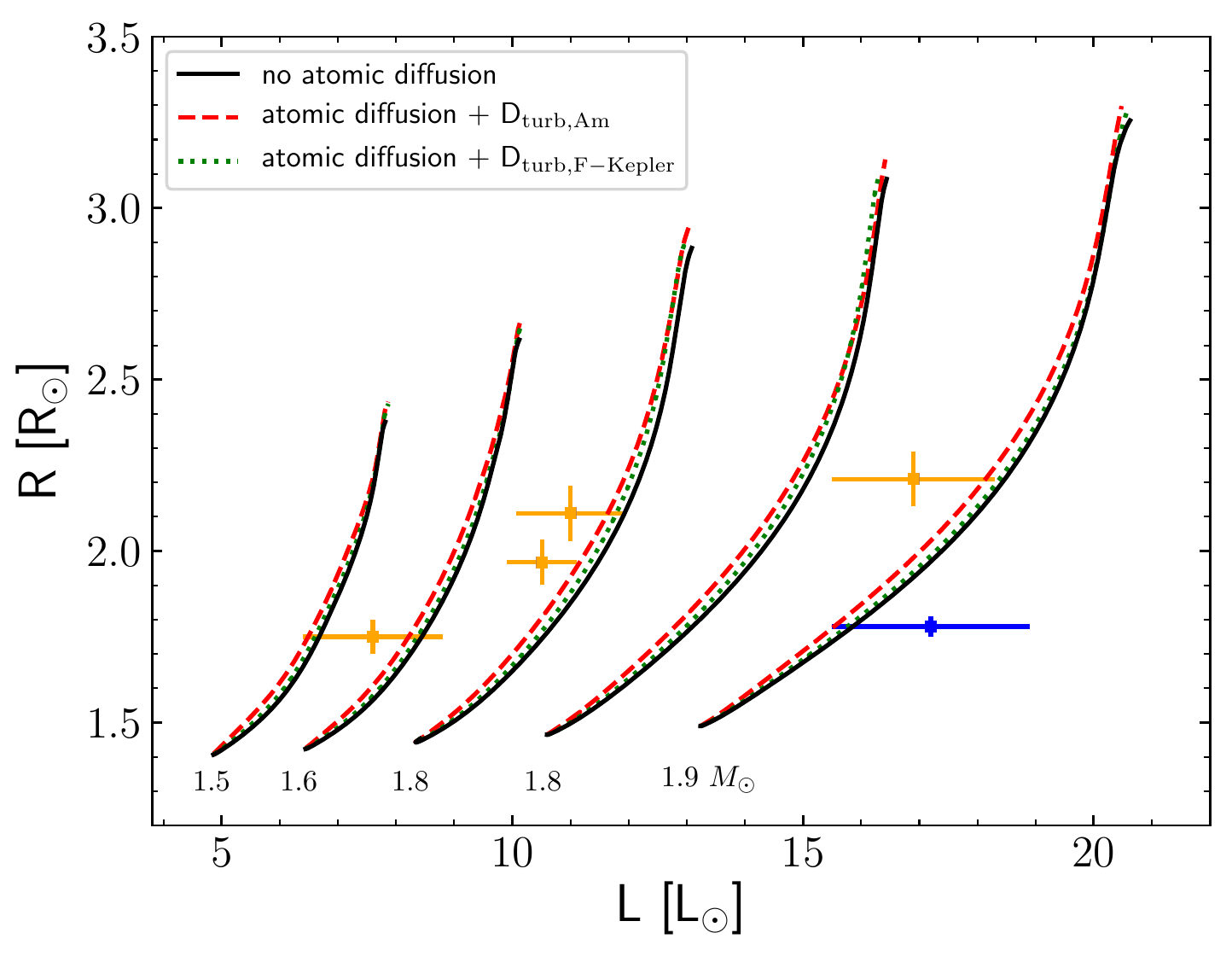}
\caption{\textit{Top}: HR diagram including evolutionary tracks for models with masses between 1.5 and 1.9~M$_\odot$ computed without atomic diffusion (black solid lines), and including atomic diffusion (with radiative accelerations) with a parametrised turbulent diffusion coefficient calibrated on Am stars (red dashed lines) and on three \textit{Kepler} F-type stars (green dotted lines). \textit{Bottom}: Radius according to luminosity for the models shown on the top panel. The blue and orange symbols are the same as in Fig.~\ref{fig:HR}.}
\label{fig:diff-rad}
\end{figure}

A-type stars have small surface convective zones and are subject to the effects of efficient atomic diffusion. Atomic diffusion is mainly the result of the competition between two processes, namely, gravitational settling, that makes elements move towards the centre of stars, and radiative accelerations, that selectively make some elements move towards the surface of stars. In stellar interiors, this competition occurs in the whole radiative zone and leads locally either to a depletion or an accumulation of each element. 

The abundance variations predicted with atomic diffusion in A-type stars (especially in Am stars) are too large compared to the observations. Hence, there is a need for a competing transport to be included in the models \citep[e.g.][]{richer00}. The efficiency of the competing transport to atomic diffusion in Am stars has been calibrated using a parametric expression first described in Eq. 1 of \cite{richer00}. When the amplitude $\omega$ and the power $n$ are fixed ($\omega=10^4$ and $n=4$) the only free parameter is the reference mass or temperature ($T_0$ or $M_0$) inside the star above which chemical composition is almost completely homogenised by the competing transport.

In order to assess the impact on the stellar properties inference from neglecting chemical element transport, we computed models including both atomic diffusion and a parametrised turbulent diffusion coefficient with the \textsc{CESTAM} evolution code. Atomic diffusion is computed as described in \cite{deal18}. The Rosseland mean opacity is computed using the OPCD package \citep{seaton05} with the improved method described in \cite{huibonhoa21}. It is important to note that despite this improved method, it is currently not possible to build a large grid of models (as the one we used in this study) incorporating these effects in a reasonable time. When a parametrised turbulent diffusion coefficient is included in the model, we use the calibration proposed by \cite{michaud11} to explain the surface abundances of Sirius A, which uses a reference mass $M_0=2\times 10^{-6}$~M$_\odot$ (hereafter $D_\mathrm{turb,Am}$). In order to test a more efficient transport (as suggested by our test of Section~\ref{result-sismo}), we also use the parametrisation calibrated on three \textit{Kepler} Legacy F-type stars \citep{verma19b}, with a reference mass of $M_0=5\times 10^{-4}$~M$_\odot$ (hereafter $D_\mathrm{turb,F-Kepler}$). Models are computed with the same input parameters as the models of the middle panel of Fig.~\ref{fig:HR}, expect for the transport of chemicals. In what follows we report the results from these tests.

\subsubsection{Impact of an internal helium gradient}

In Fig.~\ref{fig:diff}, we compare three evolutionary tracks computed without accounting for the transport of chemical elements with otherwise similar tracks, but including gravitational settling and the parametrised turbulent diffusion coefficient $D_\mathrm{turb,Am}$. Structurally, the main impact  of these transport processes is to build an internal helium gradient and to deplete heavy elements from the surface. As a consequence of this depletion, there is also a decrease of the size of the surface convective zone. We see that the main combined impact of gravitational settling and $D_\mathrm{turb,Am}$ on the stellar properties, is to slightly increase the duration of the main sequence (about $50$~Myr at maximum for the $1.5$~M$_\odot$ model). At a given age, the luminosity, effective temperature, and radius are very similar, with differences below 1\%. This indicates that to not consider the combined effect of gravitational settling and a parametrised turbulent diffusion coefficient has a negligible impact on the stellar properties inferred in this study.

\subsubsection{Impact of radiative accelerations}

Radiative accelerations play an important role in A-type stars and should be included when atomic diffusion is considered in stellar models. In Fig.~\ref{fig:diff-rad}, we compare  evolutionary tracks computed without chemical transport with otherwise similar tracks, but including atomic diffusion (with radiative acceleration) and two different parametrisations of the turbulent diffusion coefficient ($D_\mathrm{turb,Am}$ and $D_\mathrm{turb,F-Kepler}$). As shown in Section~\ref{result-sismo}, $D_\mathrm{turb,Am}$ is not sufficiently efficient to explain the observed iron abundance at the bottom of the photosphere in HD\,24712. This indicates that the reference mass of the parametrisation should be larger. We then expect the actual paramerisation of the turbulent diffusion coefficient in Ap stars to lay between $D_\mathrm{turb,Am}$ and $D_\mathrm{turb,F-Kepler}$.

In these models, iron accumulates at the bottom of the surface convective zone due to radiative acceleration. As a consequence of the efficient convective mixing, this increases the iron abundance in the convective layers. As elements accumulate where they are the main contributor to the opacity (i.e. where they absorb a lot of photons and are more subject to radiative accelerations), this iron accumulation leads to a local increase of the opacity at the bottom of the surface convective zone, hence to an increase of its size. It has the effect of slightly decreasing the effective temperature and increasing the radius of the star, as seen in the top and bottom panels of Fig.~\ref{fig:diff-rad}, respectively. The deeper the turbulent mixing, the smaller the accumulation, hence the smaller effect on the stellar properties, as seen for $D_\mathrm{turb,F-Kepler}$. As surface convective zones are smaller for larger masses, these effects are smaller at 1.9 than 1.5~M$_\odot$. The modification of the surface convective zone in the presence of radiative accelerations has already been shown in earlier works for F-type stars \citep{turcotte98b} and solar-like stars \citep{deal18}.

Inspection of the bottom panel of Fig.~\ref{fig:diff-rad} shows that the impact from the transport processes on the radii and luminosities of the stars in our sample is smaller than the observational uncertainties in these parameters. As the observed radius and luminosity were the only classical constraints considered in the modelling, the decision to not include atomic diffusion (with radiative accelerations) in the models is expected to have a negligible impact on the inferred stellar properties. The differences may even be smaller if we consider the fact that convection may be inhibited by the magnetic field, hence reducing the impact of radiative accelerations (similarly to the decreasing effect we see for larger masses).

\subsection{Impact of magnetic field}\label{discu-mag}

Recent modelling approaches\footnote{\url{https://doi.org/10.5281/zenodo.3250412} \\ \url{https://doi.org/10.5281/zenodo.3734209}} using the \textsc{mesa} software instrument have accounted for two surface effects of fossil magnetic fields in massive star models \citep{Keszthelyi19,keszthelyi20}. These approaches rely on two long-term phenomena which result from the magnetospheric-wind interaction, namely, mass-loss quenching (which reduces the mass-loss rate of the star, e.g. \citealt{ud02,bard16}) and magnetic braking (which reduces the rotation rate of the star, e.g. \citealt{ud08,ud09}). Thus far such state-of-the-art models did not cover the mass range presented in this work. To this extent, we employ this modelling approach with the goal to determine how much these effects can modify the fundamental stellar properties of Ap stars.
Of course, an important caveat is that in these models the internal magneto-hydrodynamical effects are not yet implemented for fossil fields, however, semi-analytical methods \citep{mathis05,duez10} do exist. A strong fossil magnetic field leads to other internal changes in the star, for example, by suppressing the inefficient convective regions related to opacity peaks caused by the ionisation of hydrogen and helium near the stellar surface. Nonetheless, such an effect has a negligible impact on the classical observables of an Ap star model \citep[e.g.][eqs 16 and 17]{balmforthetal01} or, more generally, on the main sequence evolution of stellar models even up to 10\,M$_\odot$ \citep[see, e.g. Figure 11 of][]{jermyn2020}.

\subsubsection{\textsc{mesa} model description}

%%
%%%% -----------------------
\begin{figure*}[!ht]
\center
\includegraphics[scale=0.18]{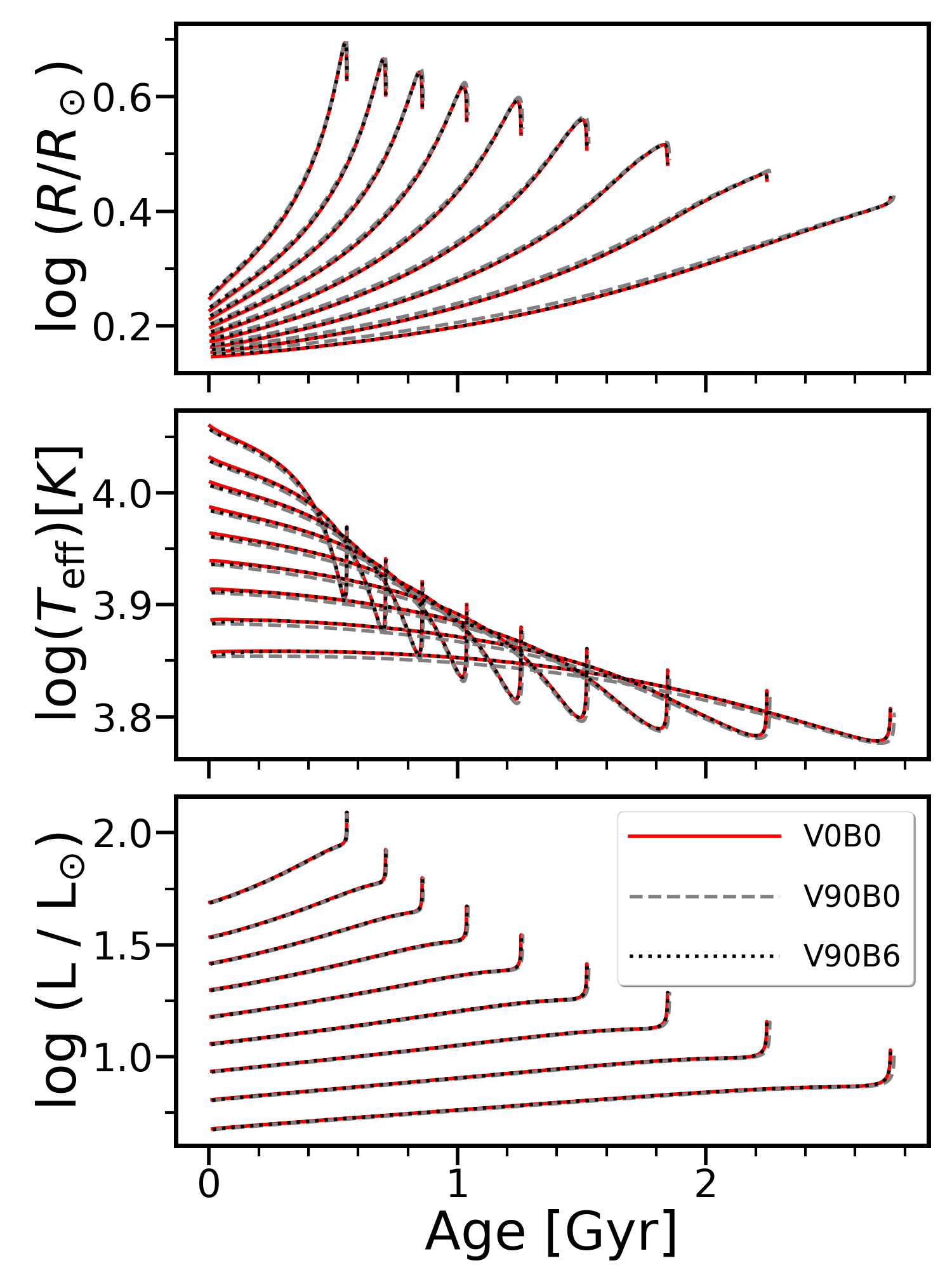}\hspace{1cm}\includegraphics[scale=0.18]{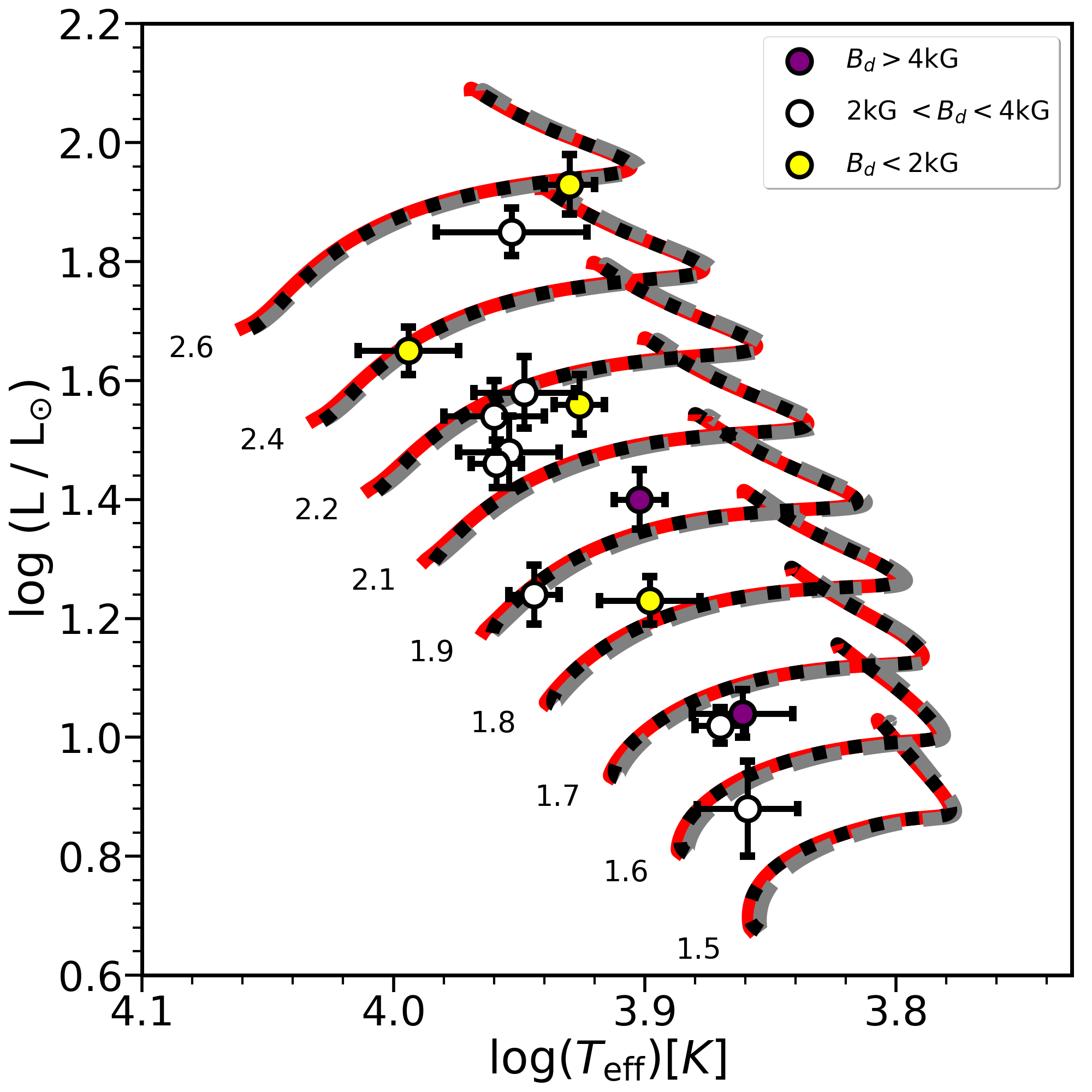}
\caption{Shown are three sets of \textsc{mesa} evolutionary models: with $v_{\rm ini} = 0$~km\,s$^{-1}$ and $B_{\rm ini} = 0$ (red solid lines), $v_{\rm ini} = 90$~km\,s$^{-1}$ and $B_{\rm ini} = 0$ (grey dashed lines), and  $v_{\rm ini} = 90$~km\,s$^{-1}$ and $B_{\rm ini} = 6$\,kG (black dotted lines). The 14 sample stars are colour-coded according to their measured magnetic field strengths as purple (> 4 kG), white (2 kG < and < 4~kG), and yellow (< 2 kG). The stellar mass in solar units is indicated next to the tracks on the right panel. In this mass range, both (moderate) rotation and surface fossil magnetic fields lead to small effects, leading to closely overlapping evolutionary tracks.}
\label{fig:mesam}
\end{figure*}
%%%% -----------------------
%%

We use \textsc{mesa} r-12115 and modelling assumptions that are similar to those of \cite{keszthelyi20}. The initial abundances are adopted as $Z_{\rm ini}=~0.014$, $Y_{\rm ini} = 0.266$, $X_{\rm ini} = 0.72$, and the initial distribution of metals follows the description of \cite{asplund09}, with isotopic ratios adopted from \cite{lodders03}. The convective mixing is adopted with $\alpha_{\rm MLT} = 1.8$ and the Ledoux criterion is used to determine the convective boundary. Overshooting is applied with the exponential method, with parameters $f= 0.021$ and $f_{0} =~0.006$, which approximately corresponds to $\alpha_{\rm ovs} = 0.15$.

For simplicity, we adopt a mass-loss rate of $10^{-14}$~M$_\odot$\,yr$^{-1}$ constant in time. Magnetic braking is modelled in the 'internal' scheme as described by \cite{keszthelyi20}, considering that all layers of the star are torqued. Chemical mixing and angular momentum transport follow the usual \textsc{mesa} methods described by \cite{paxton13}, and we do not include atomic diffusion.

We compute three sets of models in the mass range from 1.5 to 2.6~M$_\odot$. The first set of models is computed for zero rotation and zero magnetic field strength (labelled 'V0B0'), the second is for an initial rotational velocity of $v_{\rm ini}=$\,90~km\,s$^{-1}$ and zero field strength ('V90B0'), and the third is for $v_{\rm ini} =$\,90~km\,s$^{-1}$ and $B_{\rm ini} =$\,6 kG initial magnetic field strength ('V90B6'). In the latter ones, the magnetic field weakens over time, following Alfv\'en's theorem of magnetic flux conservation. Here, since we are interested in how much the surface effects of fossil magnetic fields modify the fundamental properties, we assume ZAMS values for the rotational velocity and magnetic field strength that are somewhat higher than the currently measured maximum rotational velocity and magnetic field strength in the sample. Since on the main sequence both the rotational velocity and the magnetic field strength are expected to decline over time, the initially somewhat higher values reasonably approximate the sample's mean properties at their current evolutionary stage (see Table~\ref{table:1}). Since this approach allows for testing a sort of maximum effect\footnote{If the initial rotation was higher, one would need to invoke a much stronger magnetic field to brake the rotation such that it can match the present-day rotation. On the other hand, an initially much stronger magnetic field would remain far too strong to be reconciled with the present-day values from the spectropolarimetric observations. This delicate balance is further complicated by the various evolutionary stages represented in the sample, which could only be fully self-consistently resolved on a star-by-star basis.}, the exact values are not crucial for this test.

\subsubsection{\textsc{mesa} modelling results}

Figure \ref{fig:mesam} shows the computed model diagnostics regarding their fundamental properties. In this mass range, the impact on the stellar luminosity is negligibly small (unlike higher-mass models, where complete evolutionary paths can be modified, see, e.g. \citealt{petit17,georgy17}).

The effective temperature and stellar radius differ in the case of rotating models, shifting the ZAMS location to lower $T_{\rm eff}$ and larger $R_\star$ (compared to non-rotating models) to find mechanical equilibrium. Rotating, non-magnetic models (grey dashed lines) would therefore evolve differently (quantitatively depending on the considered initial rotation), affecting the diagnostics.

The models with rotation and an initial magnetic field strength of 6 kG (black dashed lines), however, spin down due to magnetic braking. Therefore, as shown in Figure~\ref{fig:mesam}, despite the initially different ZAMS position, they converge to the non-rotating models \citep{Keszthelyi19,keszthelyi20}.

Considering various phases during the evolution, we find that the maximum difference in fundamental properties is 2~\% when comparing non-rotating, non-magnetic models with rotating magnetic models with \textsc{mesa}. This is, of course, with the important provision that the magnetic models here concern only the above-mentioned surface effects. These findings strengthen the analysis and results with the \textsc{cestam} code (where this type of surface magnetic field effects are not yet implemented), underscoring that in this mass range we do not anticipate a significant shortcoming within the scope of this paper by using non-rotating and non-magnetic models.

\section{Discussion}\label{discu}

\subsection{Comparison with \cite{kochukhov06}}

\begin{figure*}
\center
\includegraphics[scale=0.80]{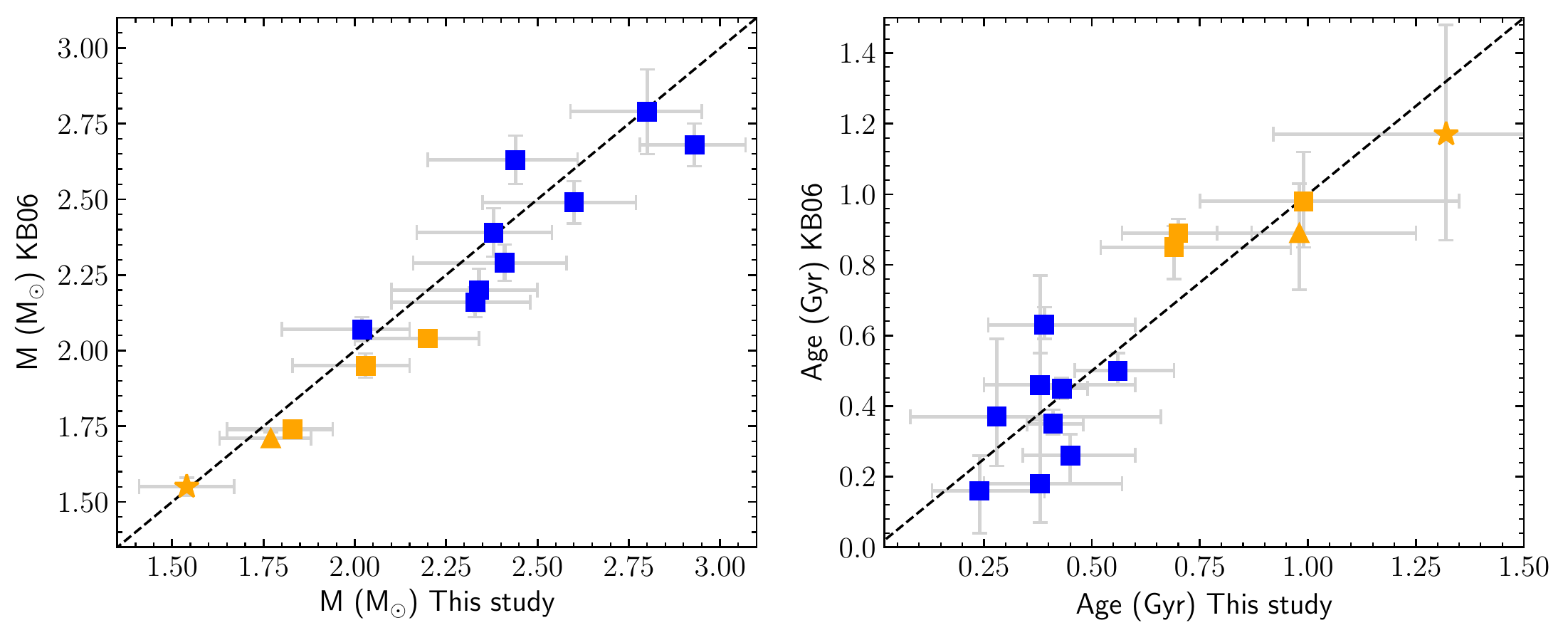}
\caption{Star-by-star comparison of the inferred mass (left panel) and age (right panel) from this study with \cite{kochukhov06}. The orange symbols are the five roAp stars of the sample while the blue ones are for the others. The star and pyramid symbols represent respectively the properties for HR\,1217 (HD\,24712) and $\alpha$~Cir (HD\,128898) inferred with the additional constraint $\Delta\nu(b)$. The dashed black lines are the 1:1 comparisons. }
\label{fig:KB06}
\end{figure*}

Fundamental stellar properties were determined for a large sample of Ap stars (including the 14 stars of this paper) by \cite[][hereafter KB06]{kochukhov06}. Masses and ages were derived using models with a unique initial metallicity $Z_\mathrm{ini}=0.018$ (in interpolating in two grids with $Z_\mathrm{ini}=0.008$ and $0.02$). As shown in Fig.~\ref{fig:KB06}, our inferences are in agreement with theirs, taking into account our $1\sigma$ intervals, except for the mass of HD\,204411 and the age of HD\,108662 which are in agreement at $2\sigma$ (between the 5$^\mathrm{th}$ and 95$^\mathrm{th}$ percentiles of the probability density functions). On the other hand, our median values are not within their $1\sigma$ intervals for nine stars out of 14. Our uncertainties are always larger. Despite the agreement within our uncertainties, we see that for ten stars out of 14, KB06 inferences underestimate the masses compare to ours. The mean systematic differences normalised by each of the errors\footnote{$1/14 \sum_{i}(M_{\mathrm{KB06},i}-M_i)/\sigma_i$} are $-0.39$ and $-1.75$ taking into account the $1\sigma$ intervals from this study (mean of the asymmetric interval) and from KB06, respectively. This comes from the fact that we considered a wider range of initial chemical compositions, and a part of the differences may also come from different input physics in stellar models (equation of state, opacity, metal mixture, convection theory, etc). For the age, the systematic difference are $-0.07$ and $0.40$, using the $1\sigma$ intervals from this study and from KB06, respectively. 

\subsection{HR\,1217 (HD\,24712): Comparison with \cite{cunha03}}

HR\,1217 (HD\,24712) was previously modelled by \cite{cunha03} using seismic constraints, taking into account different input physics and initial chemical compositions. However, the authors did not perform a complete optimisation analysis, exploring the parameter space in a systematic manner to find the best model fits and characterise the uncertainties in the inferred properties, as performed in our study. They considered masses in the range M=[1.40;1.65]~M$_\odot$, initial helium mass fraction in the range $Y_\mathrm{ini}$=[0.23;0.30], initial metal mass fraction in the range $Z_\mathrm{ini}$=[0.005;0.019], and two values of core overshoot $\alpha_\mathrm{ov}$=0.0 and 0.25. They showed that the models that were consistent with the observed large frequency separation they considered ($\Delta\nu=67.91\pm0.12~\mu$Hz) were the models with the higher values of helium ($Y_\mathrm{ini}\approx0.30$) and the smaller values of metallicities ($Z_\mathrm{ini}\approx0.005-0.009$). They also demonstrated that the mixing length parameter and core overshoot amount have less impact on the inferred stellar properties than the initial chemical composition.  

Our determination of the mass for HR\,1217 (HD\,24712), using $\Delta\nu$ with conservative uniform uncertainties ($1.54^{+0.13}_{-0.13}$~M$_\odot$) is in agreement with the mass range identified by \cite{cunha03}. Our results show that the initial metallicity at $1\sigma$ ($Z_\mathrm{ini}=0.016^{+0.009}_{-0.007}$) is slightly larger than the one of \cite{cunha03}, while we have no constraints on the initial helium content. This small discrepancy probably results mainly from the adoption of a different initial mixture of metals (\citealt{grevesse93} in their study compared to \citealt{asplund09} in our) and additional differences in the input physics.

\subsection{$\alpha$~Cir (HD\,128898): Comparison with \cite{bruntt09} and \cite{weiss20}}

$\alpha$~Cir (HD\,128898) was previously characterised by \cite{bruntt09} and \cite{weiss20}, using seismic constraints. They estimated a mass of $1.71\pm0.03$~M$_\odot$ and $1.52\pm0.15$~M$_\odot$, respectively. Our mass determination is in good agreement with that of \cite{bruntt09} and overlaps in the upper part of the $1\sigma$ interval with that of \cite{weiss20}. Similarly to \cite{cunha03}, these literature mass estimates were not performed with a complete optimisation analysis and a proper exploration of the parameter space, as presented in this paper.

\subsection{Evolutionary state of Ap stars}

\begin{figure}
\center
\includegraphics[scale=0.65]{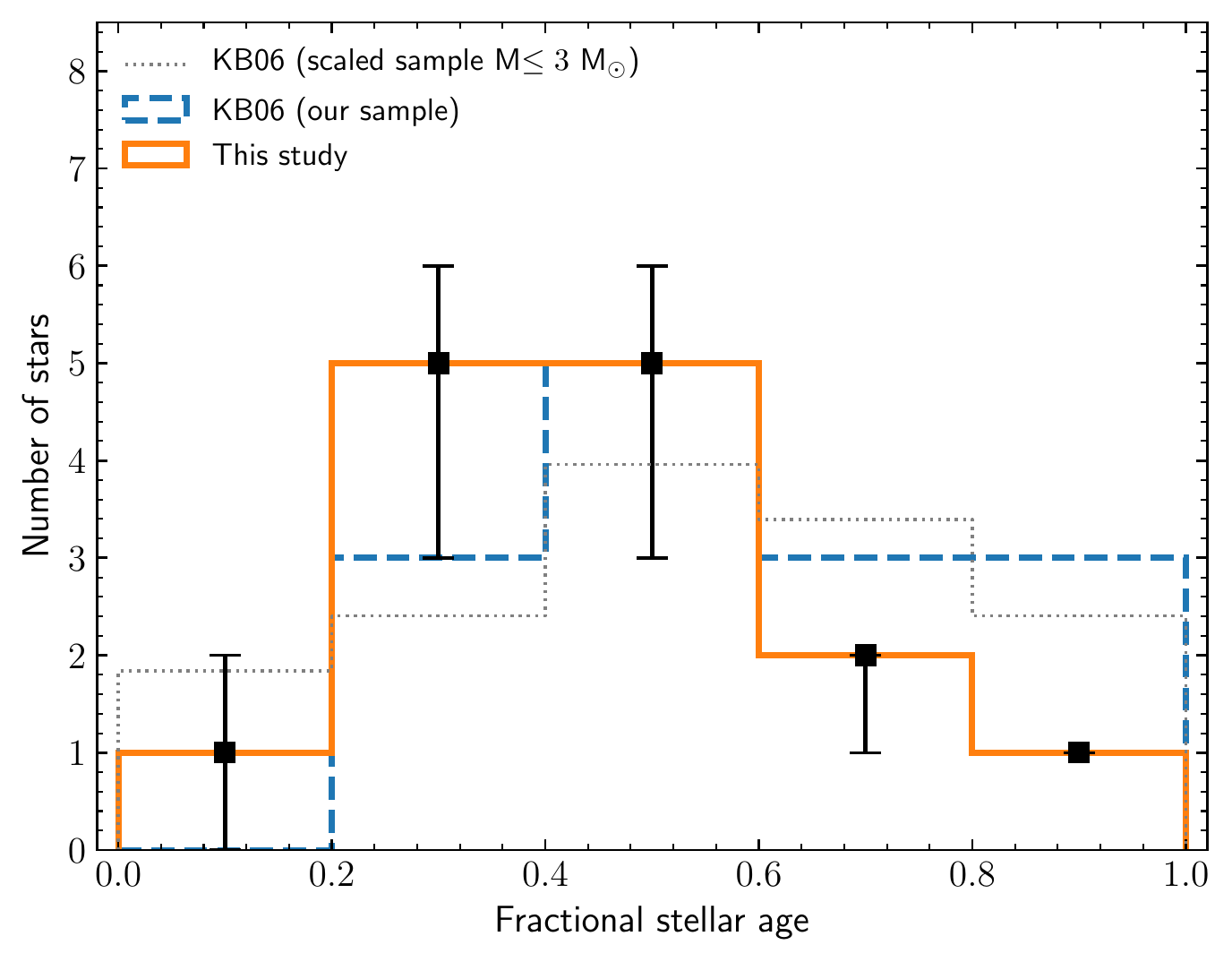}
\caption{Distribution of the relative ages of the stars in our sample according to this work (orange histogram) and to \cite{kochukhov06} (blue dashed histogram). We also show a scaled distribution from the full sample of \cite{kochukhov06} of stars with masses smaller or equal to 3~M$_\odot$ (grey dotted histogram).}
\label{fig:Xcevol}
\end{figure}

The possibility that a correlation may exist between the presence of a magnetic field and the stellar age is still a matter under debate. \cite{hubrig00} found that for stars with masses smaller than 3~M$_\odot$, magnetic fields only appear after they have spent about 30\% of their lifetime on the main sequence. Later, other studies suggested that magnetic fields could appear earlier during the main sequence \citep{bagnulo03,kochukhov06,landstreet07}. In particular, using a significantly larger and less biased sample, KB06 found that 22 stars out of 125 with M~$\le 3$~M$_\odot$ had a fractional age smaller than 0.3, rejecting the conclusion by \cite{hubrig00}. Nevertheless, for stars with M~$\le 3$~M$_\odot$, the authors found a tendency for clustering in the middle of the main sequence, which was even more evident when only stars with M~$\le 2$~M$_\odot$ were considered. Similarly, based on an ensemble study of cluster Ap stars, \cite{landstreet07} found that a significant fraction of the stars with M~$\le 3$~M$_\odot$ have small fractional ages. Figure~\ref{fig:Xcevol} shows the fractional time on the main sequence for the 14 stars in our sample, where the error bars account for the $1\sigma$ probability density functions of the initial and central hydrogen content. We find that a large fraction of the stars in the sample have already spent at least 20\% of their lifetime on the main sequence, reaching a maximum between 20 and 60\%, consistent with the clustering around the middle of the main sequence found by KB06. Moreover, we find the youngest star (HD\,220825) to have a significant probability to have spent less than 20\% of its lifetime on the main sequence, while $5^{+1}_{-2}$ other stars have completed between 20 and 40\% of their main sequence lifetime. Looking only at the stars in our sample, we find that our results predict, on average, smaller fractional times on the main sequence than those by KB06 (orange histogram compared to the dash blue one).

\subsection{roAp vs Ap stars}
The question of what drives pulsations in roAp stars is also a matter of debate \citep[e.g.][and references therein]{cunha02,cunhaetal13}. Progress in the understanding of that process requires knowledge of any systematic differences that may exist between pulsating and non-pulsating Ap stars \citep[e.g.][]{north97,hubrig00,ryabchikova04}. Despite being small, our sample has the advantage of including only stars for which the classical parameters are accurate, thus, allowing us to search for hints of potential differences between the properties of stars in these two groups. Inspection of Fig.~\ref{fig:dydz} shows that the roAp stars in our sample are systematically less massive and older than the Ap stars that are not known to pulsate. This in line with earlier findings that the roAp stars populate a narrower effective temperature range than the Ap stars in general, being located on the cooler side of the Ap stars' effective temperature distribution.

\section{Conclusion}

We inferred the stellar fundamental properties of the 14 Ap stars (including five roAp stars) characterised by \cite{perraut20}, using a grid-base modelling approach. We used the \textsc{cestam} stellar evolution code to compute the models and the \textsc{aims} optimisation method to infer the stellar properties. The grid of models included a wide range of initial chemical compositions to avoid any assumptions, and derive as unbiased as possible stellar properties. Interferometric radii and luminosities were used as classical constraints. The properties we inferred are consistent with previous work \cite[e.g.][]{kochukhov06} but with larger error bars. They are more conservative due to the wider parameter space explored, especially in the initial chemical composition. Despite the agreement, we see a trend in \cite{kochukhov06} inferences towards lower masses for stars up to 0.2~M$_\odot$. This comes from the narrower range of initial chemical composition they considered (solar metallicity).

We considered two different choices for the initial chemical composition. Firstly, we let the variation of initial helium and metallicity free, which is more suitable for a star by star analysis. Secondly, we selected only models in which the variation of these two parameters were constrained by observations, which is suitable for ensemble analyses. In both scenarios, the inferred stellar properties were very similar. We also showed that our results remain versatile and the inferred properties do not change significantly when taking reasonable assumptions and using modern implementations to model the effects of atomic diffusion, the turbulent transport of chemical elements, surface fossil magnetic fields, and stellar rotation.

Our results suggest that stars with magnetic fields can be younger than what was expected from previous studies \citep{hubrig00}. We found that between three and eight of the 14 stars have spent less than 40\% (below 20\% for HD\,220825) of their lifetime on the main sequence. Our results are in better agreement with the results of \cite{bagnulo03} and \cite{landstreet07}, that are based on cluster stars, and \cite{kochukhov06}, that is based on a large sample of field stars. Despite the fact we used a smaller sample, our study is based on observational constraints that are as accurate as one may hope to have for single stars. In addition, we find that the roAp stars in our sample have systematically lower masses and are older than the Ap stars that do not show pulsations. 

Finally, our study emphasises how knowledge of the large frequency separation $\Delta\nu$ provides an additional important constraint to the inference of the stellar properties, as illustrated by the study of the two roAp stars in our sample with known $\Delta\nu$ values (since there are five roAp stars in the sample), HR\,1217 (HD\,24712) and $\alpha$~Cir (HD\,128898). Most importantly, with the addition of $\Delta\nu$ it is possible to constrain the initial abundances. The use of seismic information thus opens new interesting possibilities to constrain the initial internal chemical composition and the transport of chemical elements in Ap stars.

\begin{acknowledgements}
MD acknowledges D. Reese for fruitful discussions and for the recent improvements of the \textsc{aims} optimisation code used for this study. This work was supported by FCT/MCTES through the research grants UIDB/04434/2020, UIDP/04434/2020 and PTDC/FIS-AST/30389/2017, and by FEDER - Fundo Europeu de Desenvolvimento Regional through COMPETE2020 - Programa Operacional Competitividade e Internacionalização (grant: POCI-01-0145-FEDER-030389). MD and MSC are supported by national funds through FCT in the form of a work contract. KP acknowledges funding from LABEX OSUG@2020 (Investissements d’avenir – ANR10 LABX56). DLH acknowledges financial support from the Science and Technology Facilities Council (STFC) via grant ST/M000877/1. We acknowledge the \textsc{mesa} developers for making their code publicly available. We thank an anonymous referee for valuable comments which helped to improve the paper.
\end{acknowledgements}  

%%%%%%%%%%%%%%%%%%%%%%%%%%%%%%%%%%%%%%%%%%%%%%%%%%%%%%%%%%%%%%%%%%%  
\bibliographystyle{aa} % style aa.bst
\bibliography{biblio.bib} % your references Yourfile.bib  
%%%%%%%%%%%%%%%%%%%%%%%%%%%%%%%%%%%%%%%%%%%%%%%%%%%%%%%%%%%%%%%%%%%
%_____________________________________________________________________

\begin{appendix}
\section{[M/H]$_\mathrm{ini}$ distribution in the grid}\label{distribFeH}

\begin{figure}[ht]
\center
\includegraphics[scale=0.6]{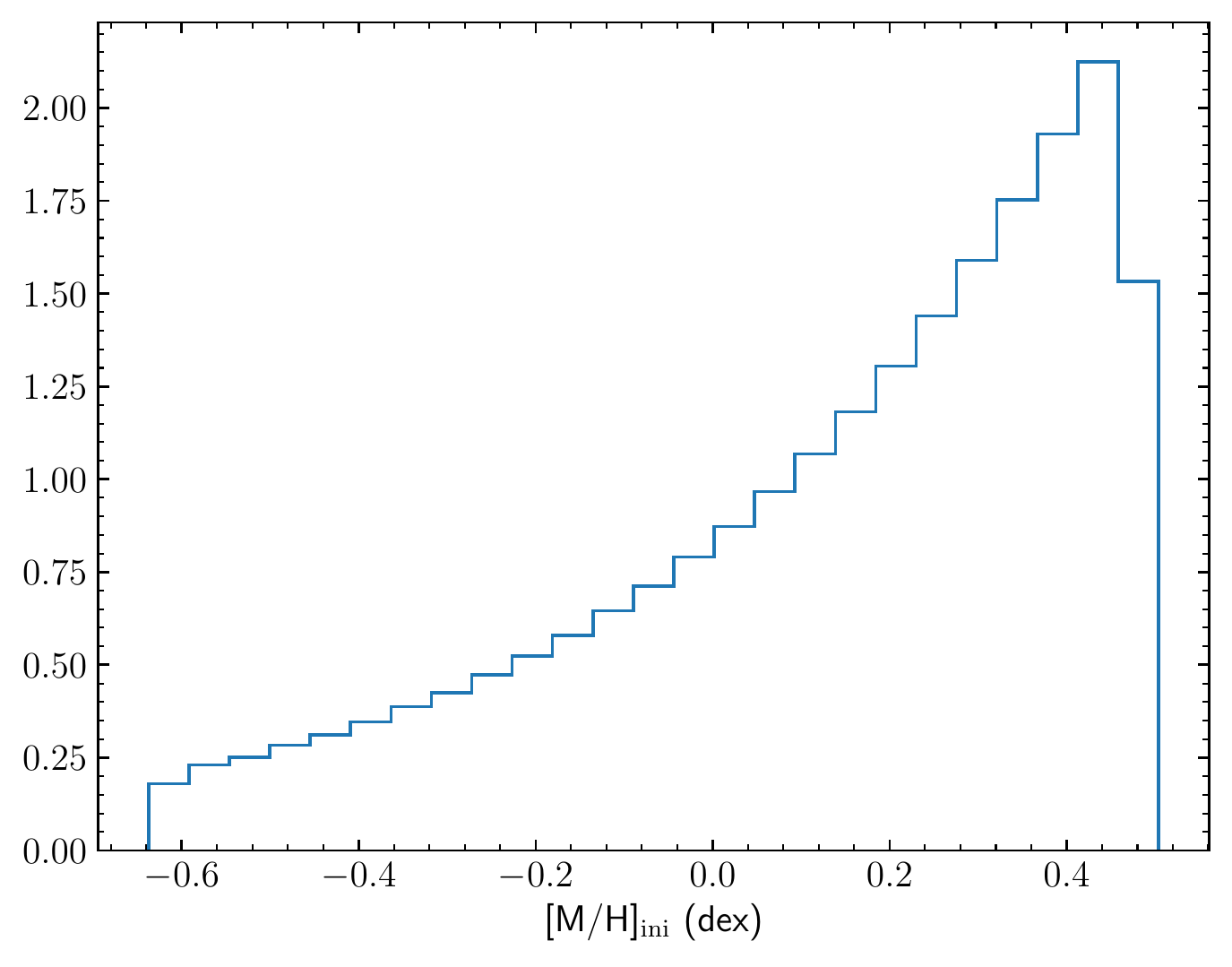}
\caption{Distribution of [M/H]$_\mathrm{ini}$ in a dense Cartesian grid (300 equally spaced values of $Y_\mathrm{ini}$ and $Z_\mathrm{ini}$) with $Y_0=[0.242;0.292]$ and $Z_\mathrm{ini}=[0.0031;0.0381]$ using an AGSS09 solar composition.}
\label{fig:distrib-FeH}
\end{figure}

Figure \ref{fig:distrib-FeH} shows the distribution of [M/H]$_\mathrm{ini}$ in a dense Cartesian grid with $Y_0=[0.242;0.292]$ and $Z_\mathrm{ini}=[0.0031;0.0381]$. 

\section{Mass, age, $X_c$, and [M/H]$_\mathrm{ini}$ distributions}\label{Appen}

Probability density functions for four parameters of the stars of the sample (except HR\,1217/HD\,24712 and $\alpha$~Cir/HD\,128898), considering the non-seismic constraints (radius and luminosity).

\begin{figure*}
\center
\includegraphics[scale=0.8]{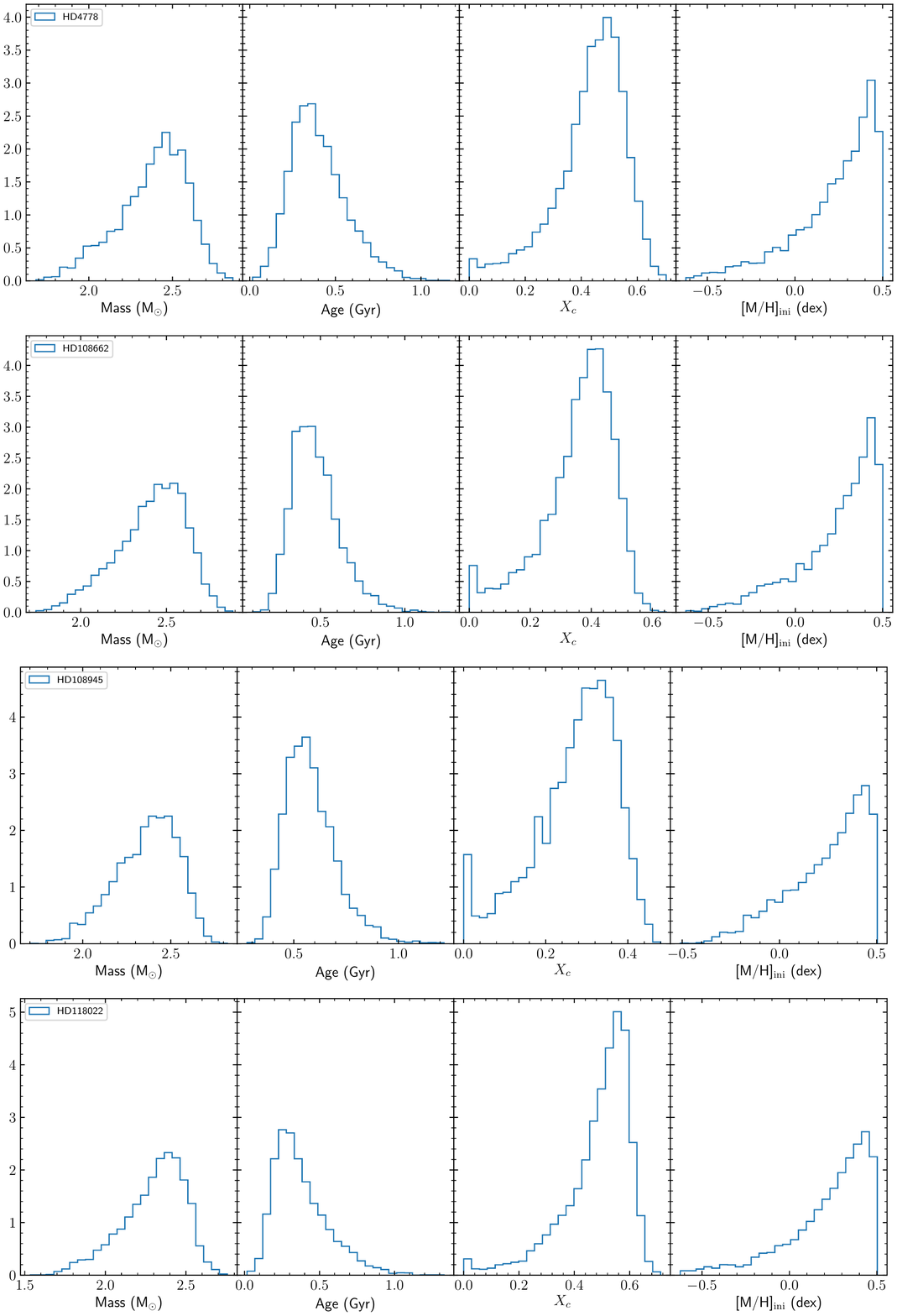}
\caption{Probability density functions for the mass, radius, hydrogen mass fraction in the core ($X_c$), and [M/H]$_\mathrm{ini}$ for HD\,4778, HD\,108662, HD\,108945, and HD\,118022.}
\label{fig:histoall1}
\end{figure*}

\begin{figure*}
\center
\includegraphics[scale=0.8]{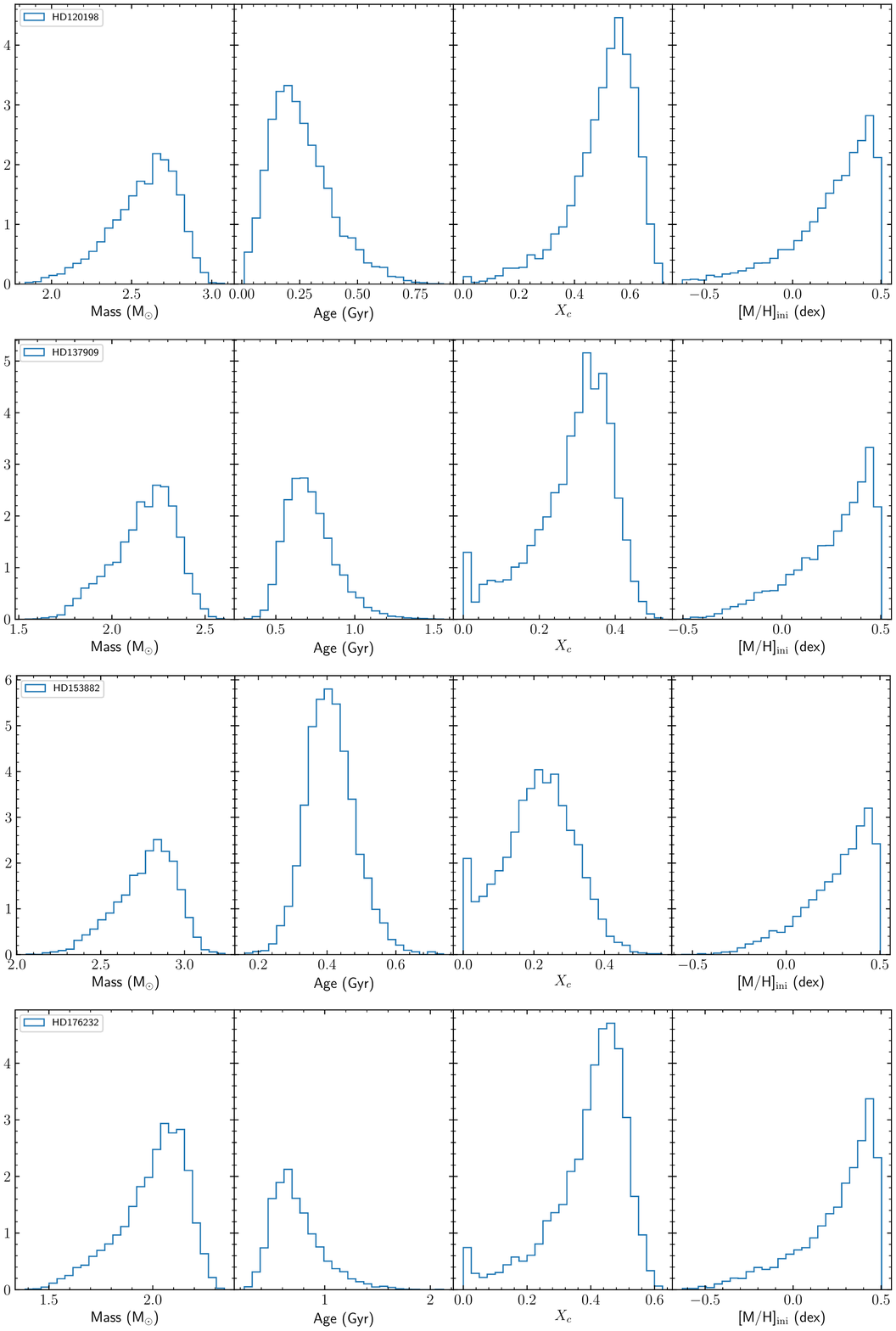}
\caption{Same as Fig. \ref{fig:histoall1}, but for HD\,12019, HD\,137909 HD\,153882, and HD\,176232.}
\label{fig:histoall2}
\end{figure*}

\begin{figure*}
\center
\includegraphics[scale=0.8]{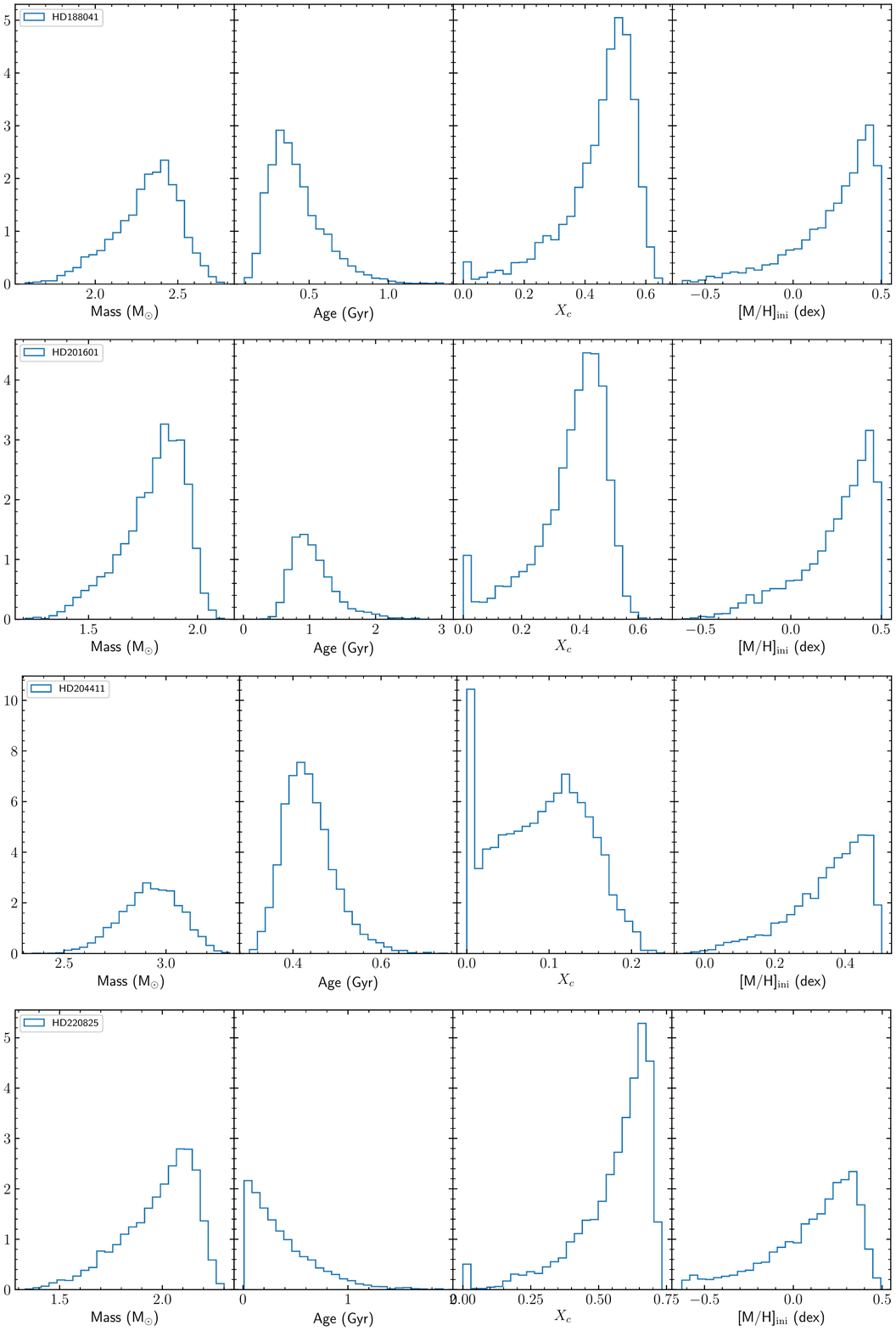}
\caption{Same as Fig. \ref{fig:histoall1}, but for HD\,188041, HD\,201601, HD\,204411, and HD\,220825 .}
\label{fig:histoall3}
\end{figure*}

\end{appendix}  
\end{document}